\documentclass[journal]{IEEEtran}
\usepackage{makeidx}
\pdfoutput=1
\usepackage[nospace,noadjust]{cite}
\usepackage{url}

\usepackage{breakurl}
\usepackage{acronym}
\usepackage{multirow}
\usepackage{adjustbox}
\usepackage{enumitem}
\usepackage{graphicx}
\usepackage{amsmath}
\usepackage{amssymb}
\usepackage{subcaption}
\usepackage{color}
\usepackage{soul}
\usepackage{listings,multicol}  
\usepackage{todonotes}
\usepackage{filecontents}
\usepackage{breqn}
\usepackage[font=sc]{caption}
\usepackage{colortbl}
\usepackage{enumitem}
\usepackage[ruled]{algorithm2e}
\usepackage{threeparttable}
\usepackage{lipsum}
\usepackage[mathlines,switch]{lineno}
\usepackage{booktabs}
\usepackage{textcomp}
\usepackage{pifont}
\usepackage{soul}
\usepackage{bm}

\acrodef{A/V}	{Audio/Video}
\acrodef{AOM}	{Alliance for Open Media}
\acrodef{AVC}	{Advanced Video Coding}
\acrodef{AWS}	{Amazon Web Services}
\acrodef{BD-BR}{Bjontegaard-Delta Bitrate}
\acrodef{bpc}	{bits per channel}
\acrodef{BQoE}	{Behavior Quality of Experience}
\acrodef{CBR}   {Constant Bitrate}
\acrodef{CABAC}	{Context-Adaptive Binary Arithmetic Coding}
\acrodef{CAVLC}	{Context-Adaptive Variable-Length Coding}
\acrodef{CDN}	{Content Delivery Network}
\acrodef{CDR}	{Continuous Dynamic Range}
\acrodef{CfE}	{Call for Evidence}
\acrodef{CIE}	{International Commission on Illumination}
\acrodef{CQoE}	{Cognitive Quality of Experience}
\acrodef{CRF}	{Constant Rate Factor}
\acrodef{CTU}	{Coding Transform Unit}
\acrodef{DASH}	{Dynamic Adaptive Streaming over HTTP}
\acrodef{DLM}	{Detail Loss Metric}
\acrodef{DR}    {Dynamic Range}
\acrodef{ECDF}  {Empirical Cumulative Distribution Function}
\acrodef{ECG}	{Electrocardiogram}
\acrodef{EEG}	{Electroencephalogram}
\acrodef{EPG}	{Electronic Program Guide}
\acrodef{EVC}   {MPEG-5 Essential Video Coding}
\acrodef{FHD}	{Full HD}
\acrodef{GOP}	{Group of Pictures}
\acrodef{GPU}	{Graphics Processing Unit}
\acrodef{HAS}	{HTTP Adaptive Streaming}
\acrodef{HD}	{High Definition}
\acrodef{HDR}	{High Dynamic Range}
\acrodef{HEVC}	{High Efficiency Video Coding}
\acrodef{HLS}	{HTTP Live Streaming}
\acrodef{HM}	{HEVC Test Model}
\acrodef{HLG}	{Hybrid Log-Gamma}
\acrodef{HLS}	{HTTP Live Streaming}
\acrodef{HVS}	{Human Visual System}
\acrodef{HW}	{Hammerstein-Wiener}
\acrodef{IP}	{Internet Protocol}
\acrodef{IPR}	{Intellectual Property Rights}
\acrodef{IQA}	{Image Quality Assessment}
\acrodef{JEM}   {Joint Exploration Test Model}
\acrodef{JND}	{Just Noticeable Difference}
\acrodef{JVET}  {Joint Video Exploration Team}
\acrodef{LF}    {Light Field}
\acrodef{LTE}	{Long Term Evolution}
\acrodef{MOS}	{Mean Opinion Score}
\acrodef{MPEG}	{Moving Pictures Expert Group}
\acrodef{MSE}   {Mean Square Error}
\acrodef{MV}	{Motion Value}
\acrodef{NARX}	{Nonlinear Autoregressive Network with Exogenous Inputs}
\acrodef{OTT}	{Over The Top}
\acrodef{OS}	{Operating System}
\acrodef{PLCC}	{Pearson Linear Correlation Coefficient}
\acrodef{PLR}	{Packet Loss Rate}
\acrodef{POP}	{Point of Presence}
\acrodef{PQ}	{Perceptual Quantizer}
\acrodef{PSNR} {Peak Signal to Noise Ratio}
\acrodef{QoE}	{Quality of Experience}
\acrodef{QoS}	{Quality of Service}
\acrodef{QP}	{Quantization Parameter}
\acrodef{RAN}	{Radio Access Network}
\acrodef{RD}	{Rate-Distortion}
\acrodef{RMSE}	{Root Mean Square Error}
\acrodef{ROI}	{Region of Interest}
\acrodef{RR}	{Reduced Reference}
\acrodef{RTMP}	{Real Time Messaging Protocol}
\acrodef{RTT}	{Round Trip Time}
\acrodef{SAO}	{Sample Adaptive Offset}
\acrodef{SC}	{Spatial Complexity}
\acrodef{SD}	{Standard Definition}
\acrodef{SDR}	{Standard Dynamic Range}
\acrodef{SI}	{Spatial Information}
\acrodef{SoP}	{Sense of Presence}
\acrodef{SQI}	{Streaming Quality Index}
\acrodef{SROCC}	{Spearman's Rank Correlation Coefficient}
\acrodef{SSIM}	{Structural Similarity}
\acrodef{STSQ}	{Short Term Subjective Quality}
\acrodef{SVC}	{Scalable Video Coding}
\acrodef{SVM}	{Support Vector Machine}
\acrodef{SVR}	{Support Vector Regression}
\acrodef{TCP}	{Transmission Control Protocol}
\acrodef{TC}	{Temporal Complexity}
\acrodef{TI}	{Temporal Information}
\acrodef{TMO}	{Tone Mapping Operator}
\acrodef{TVSQ}	{Time Varying Subjective Quality}
\acrodef{UHD}	{Ultra High Definition}
\acrodef{UDP }	{User Datagram Protocol}
\acrodef{VCI} 	{Video Complexity Index}
\acrodef{VCEG}  {Video Coding Experts Group}
\acrodef{VMAF}  {Video Multimethod Assessment Fusion}
\acrodef{VoD}	{Video on demand}
\acrodef{VQA}	{Video Quality Assessment}
\acrodef{VQM}	{Video Quality Metrics}
\acrodef{VQMT}	{Video Quality Measurement Tool}
\acrodef{VBR}	{Variable Bit Rate}
\acrodef{VIF}	{Visual Information Fidelity}
\acrodef{VIFP}	{Visual Information Fidelity - Pixel Domain}
\acrodef{VR} 	{Virtual Reality}
\acrodef{VVC}   {Versatile Video Coding}
\acrodef{WCG}	{Wide Color Gamut}

\begin{document}

\title{User Generated HDR Gaming Video Streaming: Dataset, Codec Comparison and Challenges}
\author{ Nabajeet~Barman,~\IEEEmembership{Member,~IEEE,} Maria G.~Martini,~\IEEEmembership{Senior~Member,~IEEE}
\thanks{Submitted 27 Feb 2021.}
\thanks{The authors are with Wireless Multimedia \& Networking Research Group, Kingston University London, UK. \newline
Email: \{n.barman, m.martini\}@kingston.ac.uk}}
\markboth{Preprint version of a draft currently under peer review.}
{Shell \MakeLowercase{\textit{et al.}}: Bare Demo of IEEEtran.cls for IEEE Journals}

\maketitle
\begin{abstract}
Gaming video streaming services have grown tremendously in the past few years, with higher resolutions, higher frame rates and HDR gaming videos getting increasingly adopted among the gaming community. Since gaming content as such is different from non-gaming content, it is imperative to evaluate the performance of the existing encoders to help understand the bandwidth requirements of such services, as well as further improve the compression efficiency of such encoders. Towards this end, we present in this paper GamingHDRVideoSET\footnote{ The dataset can be accessed using the link:  

https://github.com/NabajeetBarman/GamingHDRVideoSET.}, a dataset consisting of eighteen 10-bit UHD-HDR gaming videos and encoded video sequences using four different codecs, together with their objective evaluation results. The dataset is available online at [to be added after paper acceptance]. Additionally, the paper discusses the codec compression efficiency of most widely used practical encoders, i.e.,  x264 (H.264/AVC), x265 (H.265/HEVC) and libvpx (VP9), as well the recently proposed encoder libaom (AV1), on 10-bit, UHD-HDR content gaming content. Our results show that the latest compression standard AV1 results in the best compression efficiency, followed by HEVC, H.264, and VP9. 
\end{abstract}
\IEEEpeerreviewmaketitle

\section{Introduction}
Video gaming has been a prevalent and widely accepted form of entertainment, especially for the younger generation (usually 16-34 year old). 
Over the years, it has evolved and come a long way from small abstract games, such as "Super Mario" 
and Pac-Man,
to very complex and realistic games such as Battlefield, PlayerUnknown's Battlegrounds, etc. A recent survey shows that over 81\% of the Internet users reportedly have played games on at least one device \cite{GamingInsight}. Gaming video streaming, in general, can be divided into two major applications: interactive (also called cloud gaming) and passive (also called spectator gaming). With the rise of gaming and the growth of the gaming community worldwide, the latter 
has gained popularity. A recent survey in \cite{GamingInsight} shows that 34\% gamers have watched a live gaming stream in the past month with 24\% having watched an eSports tournament. The increasing popularity of such genre of entertainment has led to the rise of \ac{OTT} streaming services such as Twitch.tv and YouTube gaming, with Twitch.tv alone consisting of over 2 million monthly streamers and over 15 million daily active visitors with almost a million concurrent users making it the 4\textsuperscript{th} largest peak traffic generator in the US, just after 
the top three \ac{OTT} on-demand streaming services, Netflix, Google, and Apple.
%
\begin{figure*}[t!]
\includegraphics[width=1.0\textwidth]{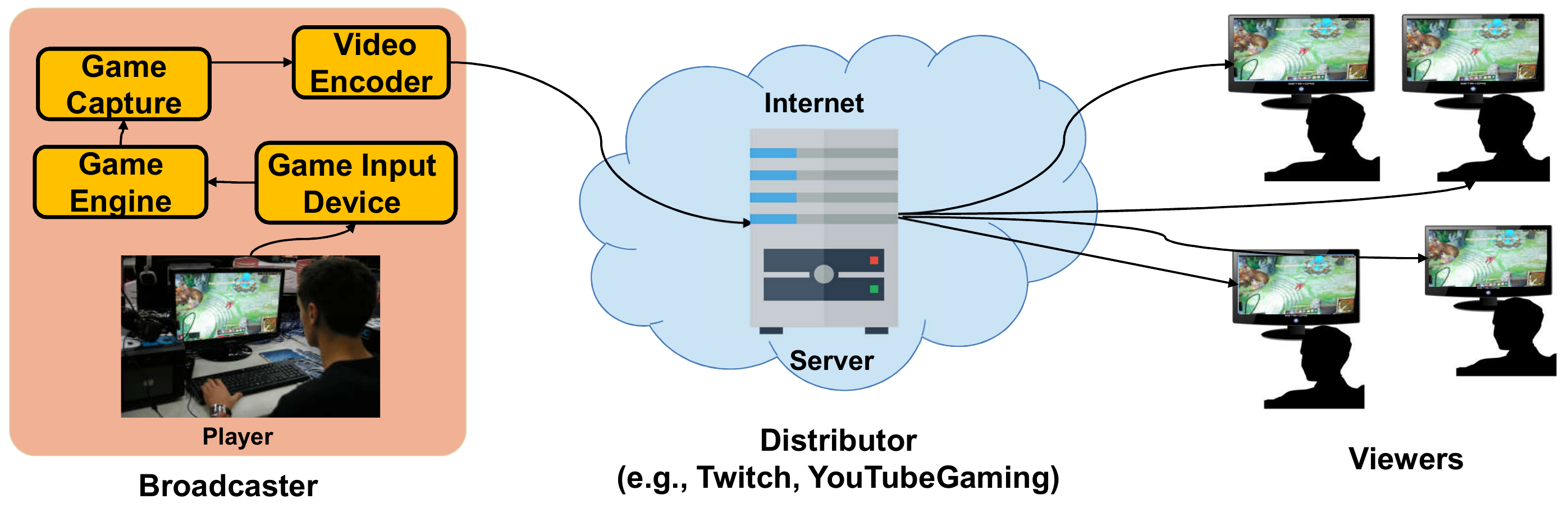}
\caption{Illustration of passive gaming video streaming applications (Spectator Gaming).}
\label{fig:SpectatorGaming}
\end{figure*}
%
Figure~\ref{fig:SpectatorGaming} shows the process of passive online gaming video streaming services over the Internet, as provided by Twitch.tv and YouTube Gaming. In such applications, the gameplay is usually performed at the broadcaster's PC. The game engine processes the game-related input commands, which are then encoded in a suitable format by the video encoders to be displayed at the end-user after decoding using the video decoder. The gameplay video and/or audio is then encoded using a video encoder and sent over the internet to the respective \ac{OTT} server, which then transcodes it into different representations and then transmits them to the end-user. Gaming 
 video streaming is challenging as well as much different from
streaming of 
traditional non-gaming content (in terms of content properties, 
performance of \ac{VQM}, streaming requirements, etc.), as discussed by 
the
authors in \cite{Barman2018QoMEX18} and \cite{Barman2018NREvaluation}.

\subsection{Motivation}

The Cisco Visual Networking Index forecasts an increase in Internet traffic, with video alone being 82\% of the net consumer Internet traffic by 2021 \cite{Cisco2017}. With the increasing introduction of newer video formats (e.g., \ac{UHD}, \ac{HDR}, \ac{LF}) and new services such as \ac{VR}, Social-TV, cloud gaming, the available network technology will not be able to meet the increased demand for high bandwidth for all the users and to satisfy users' expectations for any content, any place, any time. With the increasing availability of consumer-grade \ac{HDR} TVs, both \ac{HDR} gaming 
and
UHD-\ac{HDR} video streaming by major OTT applications such as Netflix, Amazon Prime Video, YouTube, etc., have recently caught the attention of the end-users of these applications, resulting in an increased demand for such content streaming over the Internet. 
Different strategies 
can be used to optimize the available resources 
at different stages of the
the streaming supply chain, 
such as encoding (e.g., \ac{VBR}, multiple pass, per-title encoding \cite{Netflix2015PerTitle}, dynamic optimizer \cite{Netflix2018DynamicOptimizer})), network (resource allocation, load balancing, scheduling, caching, etc.) and client (buffering, media representation adaptation, etc.). One of the widely used solutions to reduce the increasing demand for bandwidth is 
achieving high
compression efficiency during the source encoding process without loss of visual quality, thus preserving the end user \ac{QoE}. This has lead to much effort towards the development of newer, more efficient codecs such as H.265, AV1, VVC, AVS3, etc. which are shown to provide the same visual quality video encodes at a much lower bitrate than that of 
other encoders such as H.264.

\subsection{Contributions}
Given the increasing penetration and popularity of UHD-\ac{HDR} video streaming, we present in this paper a user-generated open-source UHD-\ac{HDR} gaming video dataset. Open source datasets have been an integral and very important part of research recently, especially in the field of image and video quality assessment, as they provide a shared resource for evaluation and comparison of different methods on a baseline. Following the current practices of reproducible research, open-source datasets help researchers overcome the tedious and often unnecessary process of data acquisitions, processing, and tests. Towards this end, our first significant contribution in this paper is \textit{GamingHDRVideoSET}, the first-ever
10-bit \ac{HDR} gaming video dataset, consisting of 18 gaming videos of 10 seconds duration at $3840\times2160$ resolution and 30 fps, obtained from 9 different games spanning a wide range of genres. Also, we additionally present a detailed schematic and discussion on the method to capture high-quality, 4K, HDR gaming content. To the best of author's knowledge, such a discussion on the capture methodology along with the discussion of various factors to consider while capturing 4K, HDR gaming content content is missing from the literature.

In terms of codec compression efficiency, so far, a performance evaluation considering gaming videos and real-time streaming scenarios for 10-bit, UHD \ac{HDR} gaming content is missing. It should be noted that currently almost all major gaming video streaming applications are limited to 8-bit \ac{SDR} content up to 1920$\times$1080 resolution. However, 
with the introduction of newer cloud gaming applications such as \textit{Stadia}\footnote{https://stadia.google.com/} offering 4K, \ac{HDR} gaming experience, we believe that in the very near future, the major passive gaming video streaming applications will start broadcasting also UHD/4K \ac{HDR} gaming video streaming to the viewers. 

Hence, towards this end, a second major contribution of this paper is to present an objective evaluation of the video codec compression efficiency of the four most popular video codec compression standards H.264/MPEG-AVC, H.265/MPEG-HEVC, VP9, and AV1, for live gaming video streaming applications envisioning futuristic, 10-bit UHD-\ac{HDR} gaming video streaming applications. The results of compression efficiency comparisons are reported in terms of two objective video quality metrics (\ac{PSNR} and HDR-VQM) and \ac{BD-BR} analysis, along with a discussion of frame-level quality variation. 
A discussion on the variability of bitrates between actual and target bitrate is also provided. 

It should be noted that one of the primary focuses of this paper is also on user-generated \ac{HDR} video content and the opportunities and challenges therein. Due to the importance of gaming video streaming as discussed above, we have focused here on gaming-related user-generated \ac{HDR} video content, 
but the presented dataset and learning from this work can be used for 
more generic and comprehensive user-generated \ac{HDR} image and video quality assessment research. Many different applications nowadays include both gaming and non-gaming content. Hence, the contribution of this work goes beyond considered gaming video streaming applications with the presented dataset and 
results useful in advancing the field of quality assessment and codec development and evaluation.

The rest of the paper is organized as follows. Section~\ref{sec:RelWork} presents background information about \ac{HDR} content, video quality assessment and video codecs. A discussion of video codec comparison, along with related works, is also provided. Section~\ref{sec:Dataset} 
introduces
the newly created UHD-\ac{HDR} gaming video dataset called \textit{GamingHDRVideoSET} including the game capture setup to record the games, selection criteria of the games and an analysis of the recorded video sequences in terms of 
their
spatial and temporal information and dynamic range.
Section~\ref{sec:CodecComparison} provides 
a
discussion about the selected codecs, 
the
encoding settings used as well the results of the codec comparison in terms of quality savings.
A discussion on the challenges and limitations of this work is also presented. Section~\ref{sec:Conclusion} finally concludes the paper with a discussion of future work.

\section{Background and Related Work} \label{sec:RelWork}

Considering 10-bit, UHD-\ac{HDR} gaming content, we provide first a brief discussion of \ac{SDR} vs. \ac{HDR} and 10-bit vs. 8-bit 
representation, followed by a review of the quality assessment metrics for quality evaluation. This is followed by a discussion of the video codecs 
and codec comparison,
including
the recent works comparing 
codec compression efficiency.

\subsection{\ac{SDR} vs. \ac{HDR} content}

Considering the human eye sensitivity to a wide range of luminance values present in the real world, there has been an increasing effort towards the development of capture/acquisition, processing, and display of \ac{HDR} content. Compared to \ac{SDR} content which is usually 
represented with
8 bits, \ac{HDR} content is often (but not necessarily) of higher bit-depth (usually 10 or 12) to provide 
a wider range of possible pixel values, and hence 
luminance\footnote{For a detailed analysis on the difference between 8-bit and 10-bit content, we refer the reader to \cite{barman2019analysis}.}. There are still many open challenges, from \ac{HDR} content acquisition, processing, and display, to the development of quality assessment metrics for \ac{HDR} content. Despite the many challenges, due to its promising impact on the end-user experience, recent years have seen a proliferation of \ac{HDR} content such as movies, TV series, etc. in major OTT applications. Not only limited to traditional content, \ac{HDR} has also found acceptance in gaming due to the availability of cheaper, affordable \ac{HDR} games
and
\ac{HDR} enabled displays and consoles (such as Xbox One X and Playstation 4).
\begin{table*}[htbp]
  \centering
  \caption{Summary of some 
  recent 
  works on 
  video 
  codec comparison.\\
  \small{Codecs compared, type of encoder used, evaluation methodology, quality metrics used to compare the efficiency and the focus applications, along with few observations, are presented in the table. The work presented in \cite{Barman2017QoMEX17} and this current work are the only ones which focus on Gaming Video streaming applications, whereas the others focus on natural video streaming.}}  
  \resizebox{1.0\linewidth}{!}{
    \begin{tabular}{p{2em}p{2em}p{4.785em}p{5em}p{2.5em}p{6.0em}p{4.355em}p{4.2em}p{28.07em}}
    \toprule
    \textbf{Work} & \textbf{Year} & \textbf{Codecs Compared} & \textbf{Reference/ \newline Practical Encoder} & \textbf{EM \newline (Obj/ Subj)} & \textbf{Metrics Used} & \textbf{Video Resolution} & \textbf{Focus \newline Application} & \textbf{Remarks} \\
    \midrule
    \cite{Barman2017QoMEX17} & 2017  & H.264, H.265, VP9 & Practical & Obj   & PSNR, SSIM, VIFp & upto 1080p & Live  & Codec comparison was 
    performed on gaming content. H.265 was found to be the best, followed by VP9 and H.264. Performance of VP9 vs. H.264 is found to be content dependent. \\
    \cite{Laude2018CodecComparison} & 2018  & JEM, AV1, HEVC & Reference & Obj   & PSNR  & upto 4K & on-Demand & JEM is the best, followed by HEVC and AV1, except for \textit{all-intra} mode where AV1 outperforms HM. \\
    \cite{Pinar2018CodecComparison} & 2018  & HEVC, VP9, AV1 & Practical  & Obj + Subj & PSNR, MOS & 720p & Broadcast & On 
    average, AV1 performs slightly worse than HEVC in terms of both obj and subj quality metrics. \\
    \cite{GuoNetflix2018PCS} & 2018  & H.264, H.265, VP9, AV1 & Practical  & Obj   & PSNR, VMAF & upto 1080p & on-Demand & Bitrate saving
    is calculated over optimum cross-resolution curves obtained using Dynamic Optimizer (DO) (separately for PSNR and VMAF). AV1 results in the best performance for both metrics, followed by x265, vp9 and then x264. \\
    \cite{Topiwala2018CodecComparison} & 2018  & VVC, AV1 and HEVC & Reference & Obj + Subj & PSNR, MOS & upto 4K, both 8 and 10-bit & \textit{on-Demand} & VVC is superior to both AV1 and HEVC and AV1 is superior to HEVC in terms of target bitrate encoding mode. \\
    \cite{Grois2018CodecComparison} & 2018  & AV1, JEM, VP9, HEVC & Reference & Obj   & PSNR  & upto UHD, 360 & \textit{on-Demand} & JEM performs 
    best, while HEVC performs better than AV1, which performs better than VP9. \\
    \cite{Li2019CodecComparison} & 2019  & AVC, HEVC, VP9, AVS2, AV1 & \textit{Practical}  & Subj  & MOS   & upto UHD & \textit{on-Demand} & AV1 performs the best by a sizeable margin and its performance is highly content dependent. \\
    \cite{Katsenou2019CodecComparison} & 2019  & AV1, HEVC & Reference  & Obj + Subj & PSNR, VMAF, MOS & upto UHD, 10-bit & on-Demand & Uses DO and Obj quality metrics for test sequence resolution selection followed by subjective scores to compute compression efficiency. AV1 performs better than HEVC in terms of VMAF but worse in terms of PSNR. In terms of subj ratings, there is no statistically significant difference between the two. \\
    \cite{Topiwala2017CodecComparison360} & 2019  & AV1, HEVC, JVET & Reference + Practical & Obj + Subj & SPSNR-NN, WS-PSNR, MOS & 8/10-bit 8K 360 (Spherical) Video & \textit{on-Demand} & Videos initially converted to 4K YUV 4:2:0 format before encoding. Considering reference software,
    JVET is the best followed by HEVC and then AV1. With rate control encoding mode, AV1 under-performs X.265. \\
    \cite{Topiwala2019CodecComparisonEVC} & 2019  & VVC, AV1 and EVC & Reference & Obj + Subj & PSNR, MOS & 8/10-bit, upto 4K & \textit{on-Demand} & VVC Superior to both AV1 and EVC in terms of both PSNR and MOS ratings, AV1 slightly better than EVC. \\
    \cite{DarkhanAV1} & 2020  & H.264, H.265 and AV1 & Practical & Obj + Subj & PSNR, SSIM, VMAF, MOS & upto FHD & Live & Evaluated on gaming video content using the GamingVideoSET dataset \cite{Barman2018Dataset}. AV1  better than H.264 and H.265 both objectively and subjectively\\
    This Work & 2020  & H.264, H.265, VP9 and AV1 & Practical & Obj & PSNR, HDR-VQM & 4K & Live & Evaluated on 10-bit HDR gaming content,
    AV1 results in the best compression efficiency, followed by H.265. \\
    \bottomrule
    \end{tabular}}
  \label{tab:CodecComparisonLiterature}
   \begin{tablenotes}
      \small
      \item EVC:  MPEG-5 Essential Video Coding, VVC: Versatile Video Coding, JVET: Joint Video Exploration Team, JEM: Joint Exploration Test Model. EM (Obj/Subj): Evaluation Methodology (Objective/Subjective).
        \newline
        The text in italics font is deduced based on the encoding setting used in the paper and is not explicitly mentioned in the paper. 
        \newline 
        Works in \cite{Topiwala2018CodecComparison}, \cite{Topiwala2017CodecComparison360} and \cite{Topiwala2019CodecComparisonEVC} used informal MOS scores rather than MOS scores obtained from ITU-T recommended subjective test procedure.
    \end{tablenotes}
\end{table*}
\subsection{Quality Assessment Metrics}

Over the years, there has been much work towards the development of image and video quality assessment metrics
aiming to predict the quality as 
perceived by the end-user. Most 
IQA and VQA metrics 
are developed and evaluated for 8-bit, \ac{SDR} content
and adapted to higher bit representations. 
Towards this end, we use in this work PSNR and HDR-VQM as the two objective quality metrics for quality estimation of the compressed gaming video sequences.

\subsubsection{PSNR}
\ac{PSNR} is one of the most commonly used metrics for both image and video quality assessment and is widely used for comparing codec compression efficiency, as discussed later in Section~\ref{subsec:CodecComparison}. Given a reference image (frame in case of video) $I_R$ and the corresponding distorted image, $I_D$ of size $M$x$N$, \ac{MSE} is defined as: 
\begin{equation}
    MSE = \frac{1}{MN} \sum_{i=0}^{M-1} \sum_{j=0}^{N-1} [I_R(i,j) - I_D(i,j)]^2 .
\end{equation}
PSNR in terms of decibels (dB) is then defined as:
\begin{equation}
    PSNR = 10 {log}_{10}\Big( \frac{{MAX^2_{I_{R}}}}{MSE} \Big)
\end{equation}
\noindent where  $MAX_{I_{R}}$ is the maximum possible value assumed by a pixel. 
Since our content here is 10-bit, $MAX_{I_{R}}$ is equal to 1024. The final value is computed as the average of the PSNR scores over all the frame-level scores for the video sequence. In this work, we only use the PSNR value of the Luminance component (Y), as 
commonly done in many other works.

\subsubsection{HDR-VQM}
The HDR-VQM 
video quality metric 
was 
proposed and designed 
specifically 
for \ac{HDR} video quality assessment. It uses perceptual uniform encoding, sub-band decomposition, color information, temporal information, etc. and using short-term and long-term Spatio-temporal pooling provides a prediction score for the quality of 
encoded \ac{HDR} video sequences. For the calculation of the HDR-VQM metric, we used the implementation available in \cite{HDRTools}.

\subsection{Video Codecs} \label{subsec:VodeoCodecs}

We selected H.264/AVC, H.264/HEVC, VP9, and AV1 as the four video compression codec standards for encoding compression efficiency comparison. H.264 is one of the most widely used codecs across the world, which was standardized in 2003 as a joint work by ITU-T Video Coding Experts Group (VCEG) and ISO/IEC JTC1 Moving Picture Experts Group (MPEG). Its successor H.265/HEVC, developed in 2013, is said to provide almost a 50\% bitrate saving
compared to H.264 due to better features such as bigger transform blocks, increased motion prediction modes, etc.  
However, the increased bitrate savings come at an additional cost of increased compression complexity. Initial issues with the royalties resulted in a slower acceptance of the codec by the industry. Due to the high cost and problems associated with royalties, Google released an open-source, royalty-free video codec, VP9, which was an extension of the existing VP8 codec. VP9 found wide acceptance in online streaming applications with support from all major browsers and is currently widely used in YouTube for both \ac{HDR} and \ac{SDR} content compression. 

More recently, open-source, the royalty-free AOM Video Codec 1 (AV1) encoder was developed as a joint collaborative effort by many companies as part of the Alliance for Open Media (AOM) consortium with its bitstream finalized in March 2018. Starting with VP9 tools and enhancements as the starting codec base and built along traditional block-based transform coding lines, new coding tools were integrated into the codec implementation. The codec has already seen support and implementation from major software and hardware companies leading to its wide adoption in services such as YouTube, Vimeo, and selected titles on Android devices by Netflix. Many additional efforts towards developments of next-generation codecs are currently ongoing. Joint Exploration Model (JEM) extends the underlying HEVC framework by modification of existing tools and by adding new coding tools as a joint collaborative efforts by the ITU-T Video Coding Experts Group (VCEG) and ISO Moving Picture Experts Group (MPEG) which have 
recently issued the Final Draft International Standard of the Versatile Video Coding 
(VVC\footnote{https://jvet.hhi.fraunhofer.de/}) codec, with an estimated bitrate gain around 40\% versus HEVC. 
Other ongoing efforts in this direction include the development of MPEG-5 EVC as a royalty-free codec as a competitor of AV1 and MPEG-5 Part 2 Low Complexity Enhancement Video Coding (LCEVC) (using V-Nova’s Perseus codec) 
finalized in October 2020.

\subsection{Codec Comparison} \label{subsec:CodecComparison}

Codecs comparison is a challenging issue, as the performance of the codecs depends on a lot of factors, 
such as the selected video sequences, the type of encoder (reference vs. 
practical),
the quality metrics used for comparison (objective or subjective), and the configuration of the codec (speed vs. quality, quality tuning, etc.) which in turn also depends on the focus application. Table~\ref{tab:CodecComparisonLiterature} 
presents some of the recent works related to 
comparison of the recently developed or under development video compression standards. The presented works are summarized under various categories such as codecs compared, 
type of software implementation used, evaluation methodology, 
metrics used for calculation of compression efficiency, and the type of videos used (resolution, 2D/360, etc.). Next, we discuss how each of these factors can influence the performance of the codecs.
\begin{enumerate}
    \item \textit{Reference vs. Practical Encoders}: As discussed in \cite{Barman2017QoMEX17}, reference software are 
    developed to provide complete features and produce compliant bitstreams and not necessarily to optimize the encoding process, which is left as an open 
    task
    for 
    developers. Practical encoders, on the other hand, provide
    an assessment of the compression efficiency as achieved by the codec implementations \cite{Netflix2016LargeScaleCodecComparison}. 
    
    Practical encoders, on the other hand, focus on performing compression with a level of complexity adequate for the use case and the compression efficiency is a consequence of the strategies implemented in the encoder
    
    Hence, the use of reference or practical encoders has an effect on the codec compression efficiency. 
    \item \textit{Evaluation Methodology (Objective vs. Subjective) and Metrics Used}: The metric used for the calculation of the compression efficiency 
    can also can result in different comparative performance assessment.
    As discussed earlier, PSNR is one of the most widely used metrics, as 
    also evident from the compared works presented in Table~\ref{tab:CodecComparisonLiterature}. Since PSNR does not always result in a high correlation with subjective ratings, \ac{MOS} ratings are sometimes also used for compression efficiency comparison in terms of subjective quality gain as would be perceived by the end-user. In 
    codec standardization, subjective quality tests have been introduced relatively recently, and the 50\% gain of H.265/HEVC vs. H.264/AVC was demonstrated via subjective testing \cite{Ohm2012CodecComparison}.
    \item \textit{Focus Application}: 
    The type of focus application can also result in different performance results, 
    since
    different use cases 
    have different encoding requirements. For on-Demand streaming applications, one can choose high-compression offline encoder settings, such as multiple-pass, veryslow preset, longer \ac{GOP} size, etc. as the encoding is not time-constrained. This
    enables obtaining
    a more efficiently encoded video stream, as a larger number of features of the compression standard can be explored and used in this case. On the other hand, for live streaming applications, due to the real-time encoding constraints, the video is usually encoded using a single pass, veryfast preset, resulting in a trade-off between speed and efficiency (since some complex features, in this case, need
    to be skipped). 
\end{enumerate}
\begin{figure*}[t!]
\begin{center}
\includegraphics[trim=1.5cm 4.5cm 1.2cm 4.5cm, clip=true,width=1.0\linewidth]{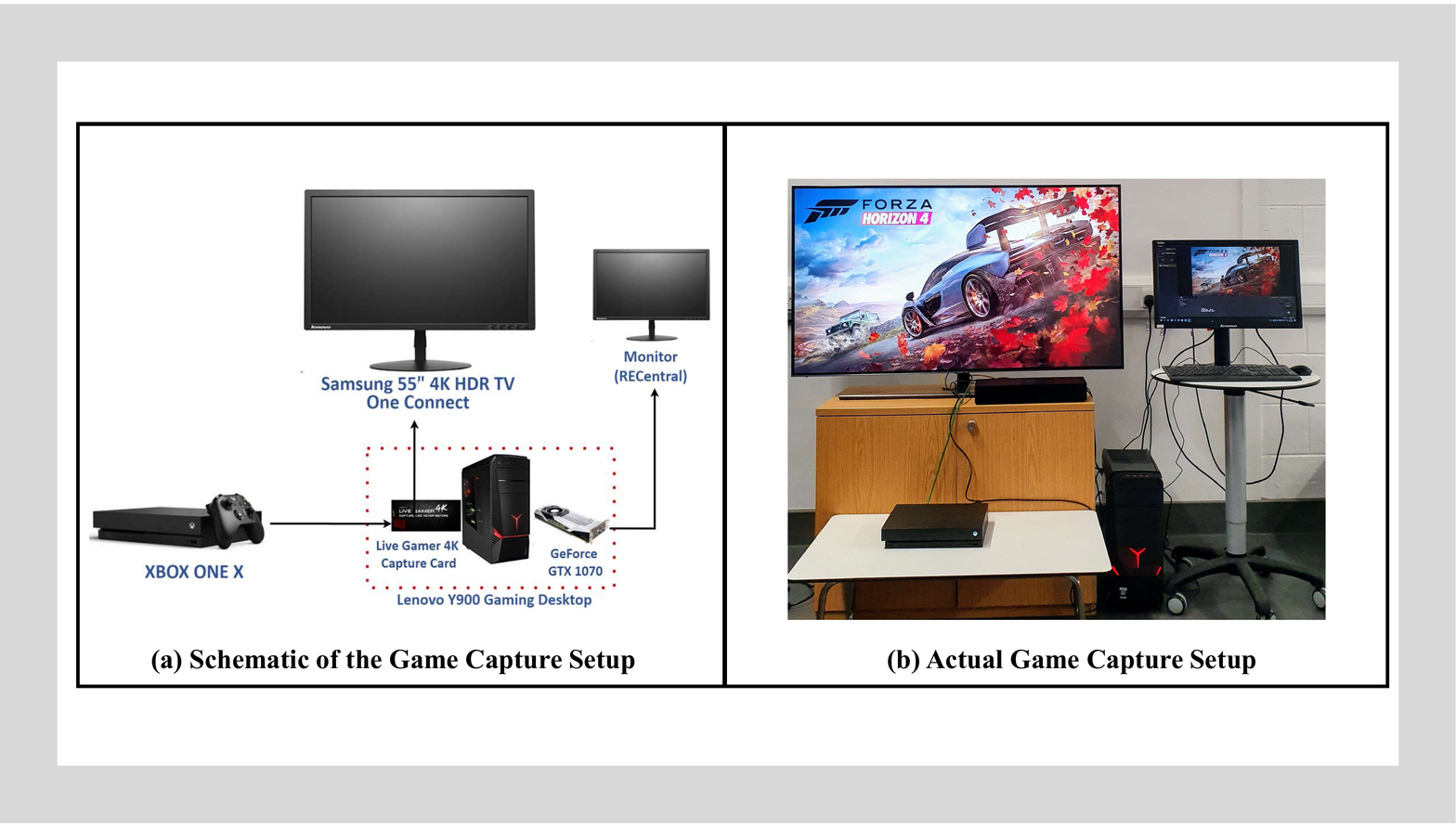}
\end{center}
\caption{Set up for capture of the 10-bit UHD-\ac{HDR} games.}
\label{fig:GameCaptureSetup}
\end{figure*}

Based on the various existing works presented here and more specifically based on the main observations mentioned under ``Remarks" in Table~\ref{tab:CodecComparisonLiterature}, it is clear that, depending on the various parameters and settings,
different works report different performance gains for AV1 compared to HEVC, VP9, and X264. Also, except for \cite{Barman2017QoMEX17}, all the works presented here are focused on on-Demand/Broadcast applications and used non-gaming 2D or 360 \ac{SDR} content. It must also be noted that some of the discussed works carried out the comparison of AV1 even before the bitstream was finalized and hence
not necessarily indicate the actual saving offered by the final AV1 codec compression standard. 

So far, there has been only one work evaluating the codec compression efficiency for gaming content using live streaming encoding settings, which is presented in \cite{Barman2017QoMEX17}. The results obtained in this work showed that the performance of the practical encoders for gaming content is highly content dependent but was limited to 8-bit, up to 1080p, \ac{SDR} content, and included only three codecs (H.264, H.265, and VP9). Towards this end, we present in this paper a codec compression evaluation of AV1, H.265, VP9, and H.264 codecs for 10-bit, UHD-\ac{HDR} gaming content using practical encoders and PSNR and HDR-VQM as the quality metrics. As discussed later in Section~\ref{subsec:EncodingSettings}, instead of using the constant QP mode of encoding, which is the more traditional, preferred method of comparison of codec compression efficiency of various compression standards, we use here instead \ac{CBR} mode of encoding. The rate-control optimization algorithm can have a big effect on the quality of the encoded video stream. Hence, the results presented here are more towards investigating the compression efficiency of the practical encoders for our application scenario rather than the core coding tools provided by the compared four compression standards.
\begin{figure*}[t!]
\begin{center}
\includegraphics[trim=1cm 2cm 2cm 1cm, clip=true,width=1.0\linewidth]{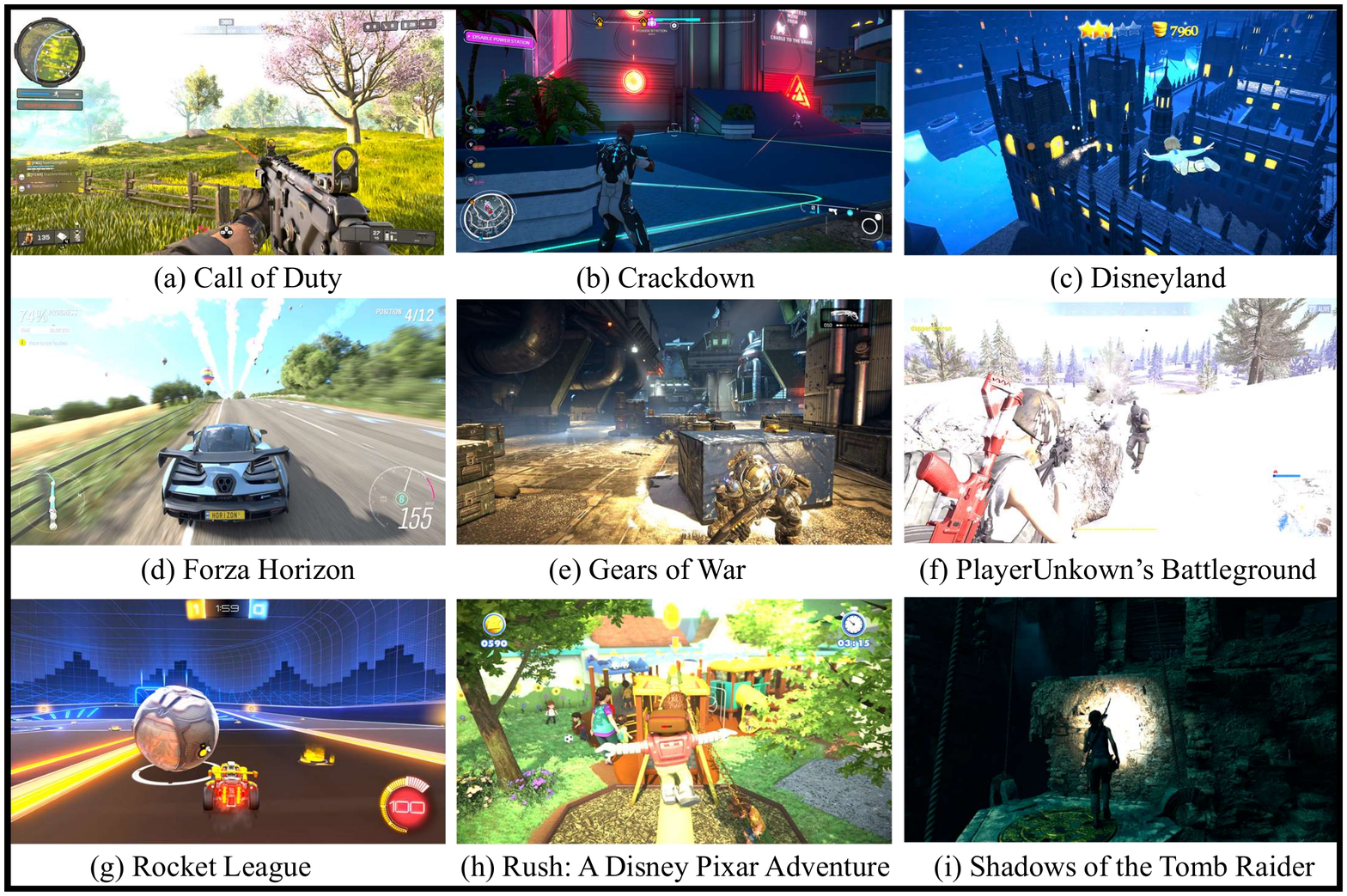}
\end{center}
\caption{Snapshots of the nine games initially considered in this work.}
\label{fig:screenshot}
\end{figure*}
\section{Dataset} \label{sec:Dataset}

As discussed earlier, an open-source dataset is of immense interest and use to the research community. So far, considering the field of quality assessment, there exists a gaming video dataset called \textit{GamingVideoSET}, presented by the authors in \cite{Barman2018Dataset}. The dataset consists of 24 high quality, 30 fps, 8-bit FHD-\ac{SDR} gaming videos obtained from 12 different games. Along with the 24 reference videos, the dataset includes subjective as well as objective evaluation results as well as compressed video sequences obtained by encoding the reference videos in 24 multiple resolution-bitrate pairs. The games were played on a FHD monitor, and the reference videos were captured using FRAPS. However, capturing 10-bit UHD-\ac{HDR} gaming videos is not that straightforward and required special hardware and software set up, which we discuss next.

\subsection{UHD-\ac{HDR} Game Capture}

In the creation of the dataset, one of the primary objectives was that the devices and configuration used are practical, not too costly (consumer-grade television, gaming desktop configuration, etc.). Figure~\ref{fig:GameCaptureSetup} shows both the setup schematic and
the actual setup used to capture and record the 10-bit UHD-\ac{HDR} gaming videos. The setup consists of a gaming console (to play the games), a capture card (to record the games), a HDR-capable monitor (to display the gameplay), a Gaming Desktop (to host the capture card and the graphics card) and a secondary monitor (to view the recording/capture software). The console output, as shown in Figure~\ref{fig:GameCaptureSetup}, is connected to the input of the capture card using a high data rate HDMI cable. The output HDMI port of the capture card is connected to the TV. The secondary monitor is connected to the graphics card of the Gaming Desktop. The recording software is running on the Gaming Desktop and is viewed on the secondary monitor.

Next, we describe each of the components we used in our actual game capture set up:
\begin{enumerate}
    \item \textit{Console:} We used the current state-of-the-art gaming console, \textit{XBox One X}, to play the games which support native UHD-\ac{HDR} gaming and offers the most premium gaming experience on a console.
    \item \textit{Capture Card:} To capture the UHD-\ac{HDR} games, we used the \textit{Live Gamer 4K - GC573} by \textit{AVerMedia}\footnote{https://www.avermedia.com/us/product-detail/GC573} which is currently the only UHD + \ac{HDR} capture card available in the market. The capture card is capable of recording gaming \ac{HDR} videos of up to UHD resolution (3840x2160) at a maximum of 60fps. 
    \item \textit{Capture Software:} To record the games, we used the \textit{RECentral} software provided by AVerMedia, which allows capturing of videos in HDR, YUV 4:2:0 chroma subsampling pixel format at a maximum bitrate of 240 Mbps encoded using HEVC codec. 
    \item \textit{\ac{HDR} TV:} Recently, \ac{HDR} TVs have increasingly found acceptance among the consumers. We used 55" Q9F Flagship QLED 4K Certified Ultra HD Premium \ac{HDR} 2000 Smart TV powered by HDR10+.
    \item \textit{Gaming Desktop:} We used Lenovo Y900 Gaming desktop with Nvidia GTX 1080 graphics card and 256GB SSD. As discussed before, the gaming desktop is required to host the capture card and also runs and records the gaming videos using the installed recording software.
\end{enumerate}
\begin{table*}[t!]
  \centering
  \caption{Description of the Games considered in this work.}
  \label{table:GamesDescription}
    \resizebox{1.0\linewidth}{!}{
    \begin{tabular}{lllp{6.93em}cp{33.145em}}
    \toprule
    \textbf{Game Name} & \multicolumn{1}{l}{\textbf{Abbr.}} &       & \multicolumn{1}{c}{\textbf{Genre}} &       & \textbf{Description} \\
    \midrule
     Call of Duty: Black Ops III & COD   &       & First Person Shooter &       & The game takes place in 2065, in a world facing upheaval from conflicts, climate change and new technologies, where one must fight to save the humanity. \\
     Crackdown 3  & CD    &       & Action-adventure &       & 
     Players explore a futuristic city, race through the streets in a transforming vehicle, and use their powerful abilities to stop a ruthless criminal empire. \\
     Disneyland & DY    &       & Open world, Action-Adventure &       & The gameplay is a combination of minigames where different players perform objectives and tasks impersonating characters. \\
     Forza Horizon 4 & FH    &       & Racing &       & A car racing game set in an open world environment based in a fictionalised Great Britain. \\
     Gears of War 4 & GoW   &       & Third-person shooter  &       & The combat game is about fighting to protect the surviving human population from decline from dangers. \\
     PlayerUnknown's Battleground & PUBG  &       & Battle royale game &       & The game is about landing on an island, looting equipment and weapons and outwitting the opponents to become the last player left standing. \\
     Rocket League & RL    &       & Sports, Football, Racing &       & A vehicular soccer game where players control a rocket-powered car and use it to hit a ball, that is much larger than the cars, towards the other team's goal area to score goals. \\
     Rush: A Disney Pixar Adventure & Rush  &       & Platform game, Action-adventure &       & A simple game where the player interacts and is taken through the worlds of different Pixar's movies such as Toy Story, Ratatouille, etc. \\
     Shadows of the Tomb Raider & SoTR  &       & Action-adventure &       & Players take on the role of Lara Croft and explores environments across the continent of Central America fighting organization and stopping an apocalypse. \\
    \bottomrule
    \end{tabular}}
    
\end{table*}
\begin{figure*}[t!]
\begin{center}
\includegraphics[width=1.0\linewidth]{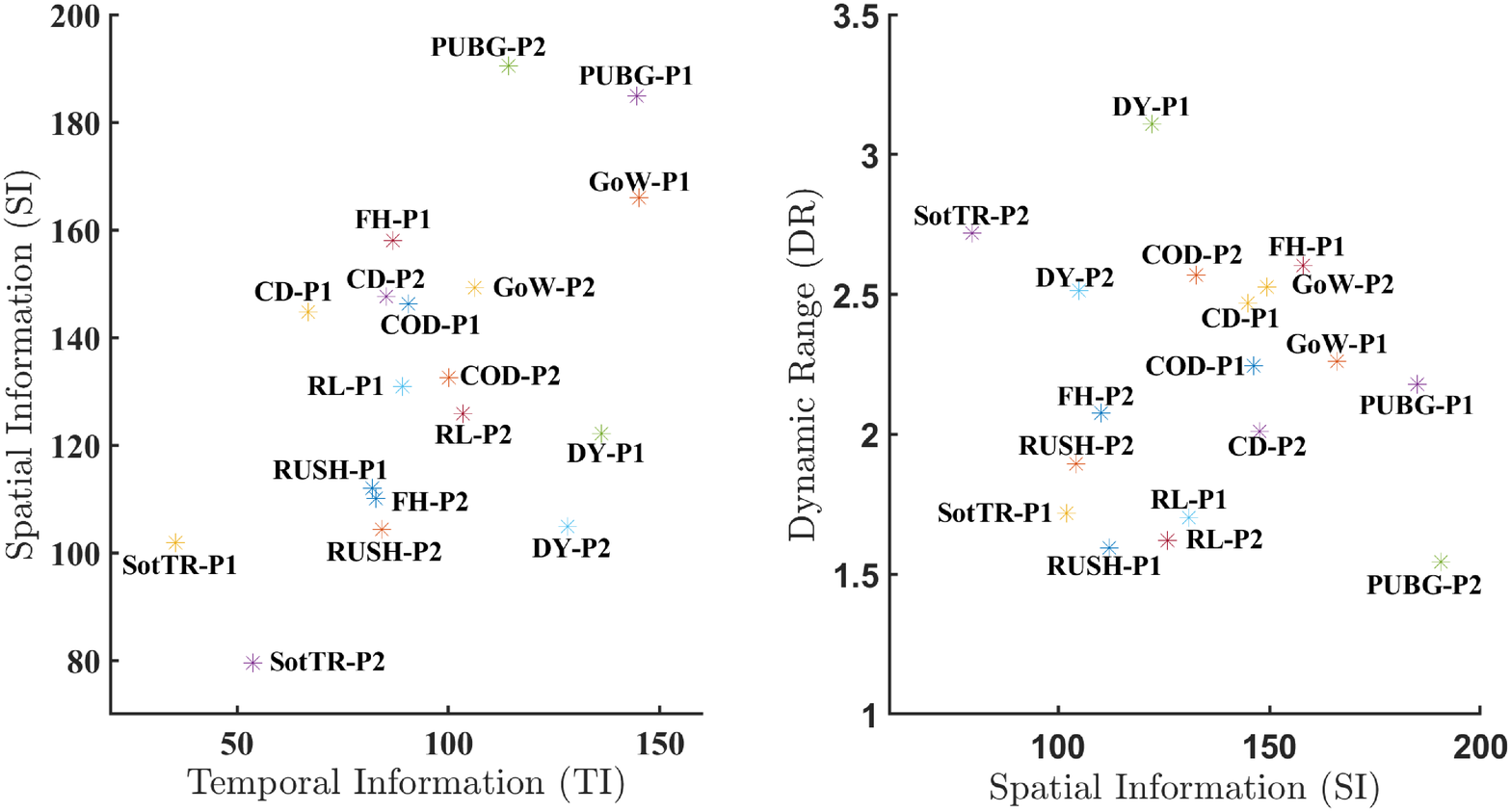}
\end{center}
\caption{\ac{SI} vs. \ac{TI} (left) and \ac{DR} vs. \ac{SI} (right) of the eighteen 10-bit reference gaming video sequences.}
\label{fig:SITI}
\end{figure*}
During the gameplay using the Xbox One X console, the above-discussed setup allows us to record gaming videos in MPEG-4 Based Media format in 10-bit YUV colorspace encoded using 
the
HEVC video codec compression standard Main profile. Before each gameplay, the \ac{HDR} picture brightness was adjusted to the requirements of the game using the inbuilt game settings as most games have settings to adjust the brightness, and sometimes other graphics settings to get the best gaming experience. In the absence of such an inbuilt game setting, the brightness was adjusted based on the best of first author's judgement. The \ac{HDR} properties of all the recorded video sequences are set to as follows:
Limited color range, BT.2020 color primaries, \ac{PQ} transfer characteristics, and BT.2020 non-constant luminance. The recorded video sequences are initially 
of various duration and 
bitrate range from 110 Mbps to 220 Mbps encoded using HEVC. Taking into account 
that 
the recorded video sequences should be representative of the actual gameplay, after a visual inspection of the video sequences
we cut two 10-second video sequences from each game using FFmpeg, removing the audio (to avoid copyright issues). The 10-bit, UHD-\ac{HDR} 10 second video sequences
are then converted into 10-bit, UHD-\ac{HDR} 10 second rawvideo (YUV, 420p chroma subsampling), which we refer to as the Reference Videos. It should be noted that though the reference videos 
obtained from the recorded video sequences are encoded, the bitrates considered for recording are high enough to not introduce any visible artifacts in the obtained reference videos which can thus be considered to be of ``reference" quality.

\subsection{Gaming Video Sequences}
The selection of the games and scenarios were performed primarily based on the following two factors (as was also done for the GamingVideoSET \cite{Barman2018Dataset}): genre and popularity of the games. Besides, we also considered the ``HDR" nature of the games,
visually inspecting various games for their \ac{HDR} picture quality 
- from very realistic games with a high dynamic range to more simplistic games, with very minimal dynamic range. Based on these factors, we selected a total of nine games, screenshots of which are shown in Figure~\ref{fig:screenshot}. Table~\ref{table:GamesDescription} presents a brief description of the twelve games along with their genres and the abbreviations used to refer to them in the rest of this paper. Based on the description and genre, it is 
clear that %
the selected games come from a wide range of genres, as would be the case for any cloud gaming or passive gaming video streaming service. The dataset and associated files can be obtained by visiting the following link: https://github.com/NabajeetBarman/GamingHDRVideoSET.

\subsection{Spatial and Temporal Information Analysis}

\ac{SI} and \ac{TI} values as defined in ITU-T Rec. P.910 \cite{ITU-P910_2008} 
are used as an approximate measurement of video content complexity. For the research study to be more realistic and inclusive, it is preferable to include sample videos from a wide range of complexity (as would be in a sample real-world application), which can be 
represented
using SI vs. TI plots. It should be noted that the SI and TI 
values
for 10-bit videos 
are higher 
than those for 8-bit videos with similar content 
(due to increased bit-depth, usually of magnitude four times higher), as discussed by the authors in \cite{barman2019analysis}. Figure~\ref{fig:SITI} 
(left) 
shows the SI vs. TI plot 
for the 18 reference video sequences. Based on the figure, it is clear that the selected video sequences do cover a wide range of complexity - from low to medium to high. It needs to be noted that various factors such as encoding, bit-depth, resolution, etc. might affect the SI and TI values as briefly discussed in \cite{barman2019analysis}, and hence the SI vs. TI plots are usually limited to reference video sequences.

\subsection{Dynamic Range of the Sequences}

Figure~\ref{fig:SITI} (right) shows the plot of \ac{DR} vs. SI of the 18 video sequences to help understand the \ac{HDR} characteristics of the reference video sequences. The dynamic range was calculated 
removing the top 1\% and lowest 1\% luminance values, i.e.,
by using the following procedure:



\begin{enumerate}
    \item Read the first frame of the video.
    \item Sort the pixel values of the luminance (Y) component of the frame in ascending order.
    \item Given that $M \times N$ is the number of pixels in the frame, calculate $log_{2}(L_{max}/L_{min})$, 
    where $L_{min}$ is the  $(M \times N/100)^{th} $ luminance value in the ordered array and $L_{max}$ is the  $((M \times N)-((M \times N)/100))^{th}$ luminance value in the ordered array.
    \item Store the value in an array.
    \item Read the next frame and repeat steps from (2) - (4) till the last frame of the video.
\end{enumerate}

The \acl{DR}
value for the video is then calculated as the average of the individual frame level \ac{DR} values stored in the array.From the figure, it is clear that the reference videos used in this study are of different dynamic range values and have a wide range of coding complexity. The DR values range approximately between 1.5 to 3.1 with a median DR value of 2.2. The wide range of DR values of the recorded gaming video sequences is a good representative of commercial-grade TV display content and can help us understand the challenges associated with processing and displaying such videos on displays with different settings.


\section{Codec Comparison} \label{sec:CodecComparison}

\subsection{Encoding Settings} \label{subsec:EncodingSettings}
\begin{table}[t!]
  \centering
  \caption{Encoder settings summary.}
  \label{table:EncoderSettings}
  \resizebox{1.0\linewidth}{!}{
    \begin{tabular}{ll}
    \toprule
    \multicolumn{1}{c}{\textbf{Parameter}} & \multicolumn{1}{c}{\textbf{Value}} \\
    \midrule
    Duration & 10 secs \\
    Resolution & 3840x2160 \\
    Frame Rate & 30 \\
    Bit-depth & 10 \\
    Pixel Format & 4:2:0 \\
    Colorspace & 
    BT.2020 NCL \\
    Color Primaries & 
    BT.2020\\
    Color Range & MPEG/Studio/TV \\
    Color TRC & SMPTE 2084 PQ \\
    Encoder & FFmpeg (v. 4.2.1-static) \\
    Video Compression Standards & H.264, H.265, VP9, AV1 \\
    Encoding Mode & Constant Bitrate \\
    Encoding Bitrates (Mbps) & 6, 12, 18, 24 \\
    \bottomrule
    \end{tabular}}
    
\end{table}
Since our focus is on user-generated \ac{HDR} content and given the fact that there currently does not exist any platform streaming UHD-\ac{HDR} 10-bit gaming content, we used the YouTube \ac{HDR} encoding blog \cite{YouTubeHDREncoding} and Twitch Broadcasting Requirements \cite{TwitchCBR} as the baseline for the choice of encoding settings as summarized in Table~\ref{table:EncoderSettings}. The encoding mode used in this work is \ac{CBR}, unlike the more widely used 
2-Pass
Average Bitrate mode of encoding as commonly used by 
on-demand video streaming applications. The reason behind this is that the user gameplay can sometimes have long, dull moments followed by very high action scenes. In such cases, the variable/average bitrate mode of encoding might result in sudden peaks in bandwidth demand and hence result in increased stalling events at the end-user \cite{TwitchCBR}.

While in real-world applications
multiple
resolution-bitrate pairs are used for quality adaptation, for brevity
here we have limited our analysis to a single resolution (UHD) and four different bitrate values, namely 
6, 12, 18 and 24 Mbps.
We encoded the reference video sequences using the four video compression standards
AVC/H.264, HEVC/H.265, VP9, and AV1, using their implementation libraries available in FFmpeg (\textit{libx264}, \textit{libx265}, \textit{libvpx-vp9} and \textit{libaom-av1} respectively). All processing was done on an \textit{Ubuntu 18.04.01 LTS} system with 16 GB RAM and 256 GB SSD + 16 TB HDD. As discussed earlier, encoding settings used have a huge impact on the obtained results, 
hence it is preferred that they are kept 
as 
similar 
as possible among the compared codecs, in order to have fair, comparable results. While there may exist different implementations for each compression standard whose performance might be (slightly) different than that obtained using the respective \textit{FFmpeg} libraries, using the same implementation allows us to make sure that the relevant settings are more or less consistent between the different encoders leading to a somewhat fair comparison for our focus application. Table~\ref{table:CodecSettings} presents the encoding settings used for the four codecs. 

Since we are not using the optimized codec implementations for each compression standard, as well as considering the fact that the used open-source AV1 codec implementation is still to be optimized for speed, in this work
encoding duration comparison is not presented. Nevertheless, in line with observations reported in our earlier work in \cite{Barman2017QoMEX17} using 8-bit, \ac{SDR} gaming content, H.264 is the fastest followed by H.265 and VP9, 
while AV1 
is much slower than the first three. It should also be noted that 
the recorded videos were already encoded using the H.265 compression standard, making this study basically an evaluation of transcoding efficiency of the video codecs, not performed in any other earlier work discussed above.
\begin{table}[t!]
  \centering
  \caption{FFmpeg codec settings summary.}
   \label{table:CodecSettings}
     \resizebox{1.0\linewidth}{!}{
    \begin{tabular}{p{6.645em}p{18.215em}}
    \toprule
    \multicolumn{1}{l}{\textbf{Encoder}} & \multicolumn{1}{l}{\textbf{Settings}} \\
    \midrule
    \multicolumn{1}{l}{libx264, libx265} & preset=veryfast, profile=main, level=4.0 \\
    \midrule
    \multicolumn{1}{l}{libvpx-vp9} & \multicolumn{1}{l}{deadline=realtime, quality=realtime, profile=2} \\
    \midrule
    \multicolumn{1}{l}{libaom-av1} & \multicolumn{1}{l}{cpu-used=8} \\
    \midrule
    libx264, libx265, \newline libvpx-vp9, libaom-av1 & single pass, buffer=bitrate, \newline closed gop=60 (2s), CBR, \newline pix\_fmt=yuv420p10le, color\_primaries=9, \newline color\_trc=16, colorspace=9, color\_range=1 \\
    \bottomrule
    \end{tabular}}    
\end{table}

\subsection{Compression Efficiency}

\begin{figure*}[t!]
\begin{center}
\includegraphics[trim=2cm 9.5cm 2cm 1cm, clip=true, width=1.0\linewidth]{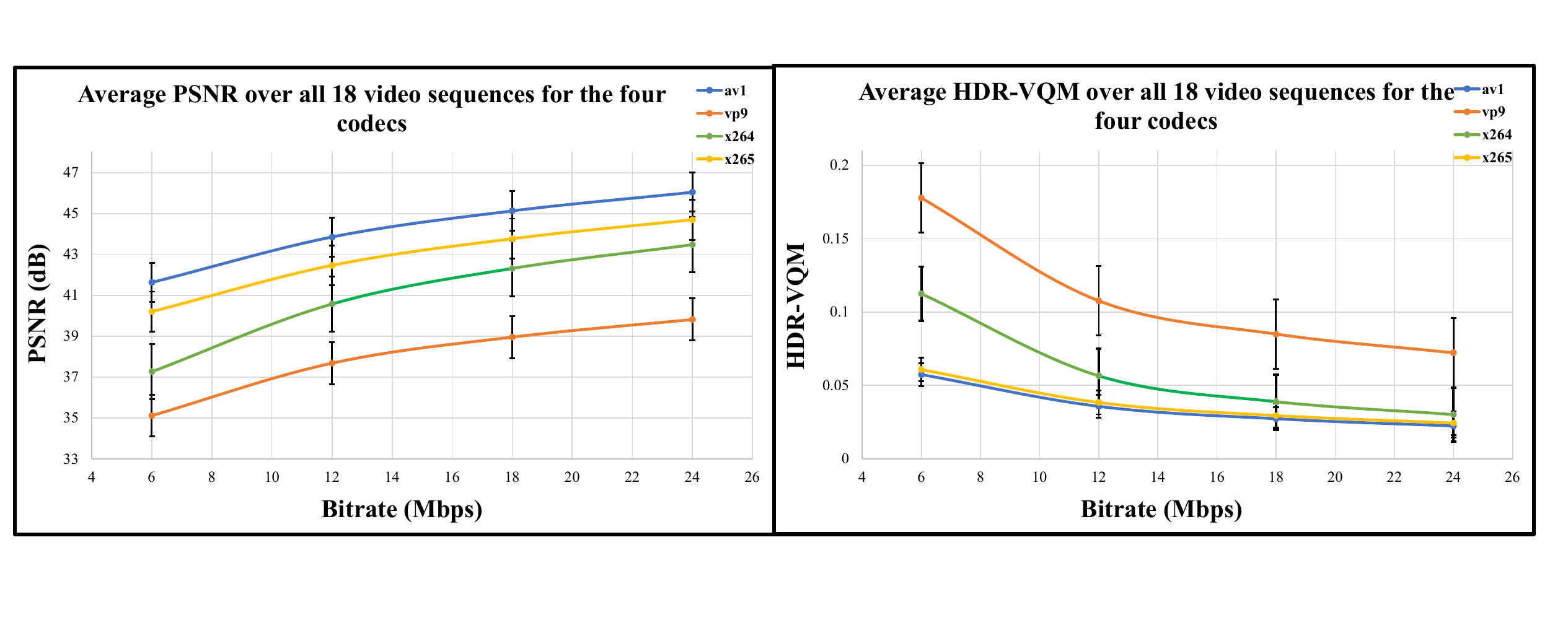}
\end{center} 
\caption{Quality vs. Bitrate plot for the four codecs in terms of PSNR (left) and HDR-VQM (right) averaged over all the 18 video sequences.}
\label{fig:CodecCompression}
\end{figure*}

We evaluated the performance of the compression standards using two full-reference \acl{VQM}: PSNR and HDR-VQM discussed earlier in Section~\ref{sec:RelWork}. Figure~\ref{fig:CodecCompression} presents the plots of 
PSNR vs. Bitrate and HDR-VQM vs. Bitrate considering the average over all the eighteen encoded video sequences for the respective bitrate. The bars represent a 95\% confidence interval for each corresponding metric and bitrate value. Based on the figure, it can be concluded that, for 10-bit UHD-\ac{HDR} gaming content, considering both PSNR and HDR-VQM metric, AV1 results in the best compression efficiency followed by HEVC, H.264, and VP9. A similar trend for H.264, H.265, and VP9 for 8-bit \ac{SDR} gaming video content was reported in our earlier work presented in \cite{Barman2017QoMEX17}. The relatively low performance of VP9 compared to H.264 is not surprising as our earlier work in \cite{Barman2017QoMEX17} reported decreased performance of VP9 compared to that of H.264 for 8-bit SDR, 1920$\times$1080, and lower resolutions. In terms of the HDR-VQM metric, AV1 performs the best but is closely followed by HEVC (as compared to PSNR, where the gap was quite noticeable). The gap between the performance of the codecs decreases
at higher bitrates.

\subsubsection{\ac{BD-BR} Analysis}

In order to quantify the amount of bitrate savings, we use \ac{BD-BR} \cite{BD-BR} analysis using PSNR. BD-BR analysis is widely used in the video quality research community to calculate the average gain in quality metric or the average percentage of bitrate savings between two Rate-Distortion curves. For more details on BD-BR computation, we refer the reader to the original publication in \cite{BD-BR}. For this work, we used the MATLAB implementation, which has been verified for accuracy based on the results presented in \cite{BD-BR} and cross-checked with the open-source implementation provided by Netflix in \cite{NetflixVMAF_Github}. Since the actual bitrate values of the encoded video are different from the target bitrate (those mentioned as an input parameter during the encoding), we extracted the actual encoded bitrate of the videos using \textit{FFprobe}\footnote{https://ffmpeg.org/ffprobe.html}. In the reported BD-BR performances, the actual bitrate values are used for calculations.
\begin{table*}[htb!]
    \centering
    \caption{BD-BR analysis results in terms of PSNR savings.}
    \label{table:BDBRresults}
    \resizebox{1.0\linewidth}{!}{
    \begin{tabular}{lcccccc}
    \multicolumn{1}{l|}{\textbf{Sequence}} & \multicolumn{1}{l}{\textbf{AV1 vs VP9}} & \multicolumn{1}{l}{\textbf{AV1 vs X264}} & \multicolumn{1}{l}{\textbf{AV1 vs X265}} & \multicolumn{1}{l}{\textbf{X264 vs VP9}} & \multicolumn{1}{l}{\textbf{X265 vs VP9}} & \multicolumn{1}{l}{\textbf{X265 vs X264}} \\
    \midrule
    \multicolumn{1}{l|}{\textbf{COD-P1}} & -6.39 & -3.14 & -0.73 & -3.28 & -5.66 & -2.37 \\
    \multicolumn{1}{l|}{\textbf{COD-P2}} & -5.95 & -2.74 & -0.63 & -3.25 & -5.32 & -2.07 \\
    \multicolumn{1}{l|}{\textbf{CD-P1}} & -6.74 & -3.64 & -1.65 & -3.13 & -5.09 & -1.96 \\
    \multicolumn{1}{l|}{\textbf{CD-P2}} & -7.62 & -3.73 & -1.74 & -3.92 & -5.87 & -1.97 \\
    \multicolumn{1}{l|}{\textbf{DY-P1}} & -8.56 & -4.49 & -2.66 & -4.09 & -5.87 & -1.77 \\
    \multicolumn{1}{l|}{\textbf{DY-P2}} & -6.72 & -3.97 & -2.04 & -2.79 & -4.68 & -1.84 \\
    \multicolumn{1}{l|}{\textbf{FH-P1}} & -5.64 & -3.32 & -1.35 & -2.33 & -4.28 & -1.93 \\
    \multicolumn{1}{l|}{\textbf{FH-P2}} & -5.80 & -3.42 & -1.47 & -2.42 & -4.34 & -1.89 \\
    \multicolumn{1}{l|}{\textbf{GoW-P1}} & -5.77 & -3.11 & -1.22 & -2.70 & -4.55 & -1.87 \\
    \multicolumn{1}{l|}{\textbf{GoW-P2}} & -6.27 & -3.22 & -1.44 & -3.08 & -4.83 & -1.74 \\
    \multicolumn{1}{l|}{\textbf{PUBG-P1}} & -4.52 & -2.00 & -1.49 & -2.54 & -3.03 & -0.50 \\
    \multicolumn{1}{l|}{\textbf{PUBG-P2}} & -4.27 & -1.99 & -1.35 & -2.30 & -2.92 & -0.63 \\
    \multicolumn{1}{l|}{\textbf{RL-P1}} & -6.22 & -3.73 & -2.04 & -2.52 & -4.18 & -1.64 \\
    \multicolumn{1}{l|}{\textbf{RL-P2}} & -6.42 & -3.71 & -2.02 & -2.72 & -4.39 & -1.66 \\
    \multicolumn{1}{l|}{\textbf{RUSH-P1}} & -5.88 & -3.33 & -1.78 & -2.58 & -4.10 & -1.50 \\
    \multicolumn{1}{l|}{\textbf{RUSH-P2}} & -7.99 & -4.71 & -2.27 & -3.33 & -5.72 & -2.38 \\
    \multicolumn{1}{l|}{\textbf{SoTR-P1}} & -6.60 & -3.02 & -1.12 & -3.62 & -5.48 & -1.85 \\
    \multicolumn{1}{l|}{\textbf{SoTR-P2}} & -6.60 & -3.38 & -1.45 & -3.26 & -5.16 & -1.88 \\
    \midrule
    \textbf{Average} & \textbf{-6.33} & \textbf{-3.37} & \textbf{-1.58} & \textbf{-2.99} & \textbf{-4.75} & \textbf{-1.75} \\
    \bottomrule
    \end{tabular}}    
\end{table*}

Table~\ref{table:BDBRresults} presents the results of BD-BR analysis in terms of gain in quality (PSNR) for the six different codec comparison combinations for all the eighteen video sequences and also the average across the eighteen video sequences. Similar observations are obtained when using HDR-VQM as the quality metric instead of PSNR, which is provided in the GamingHDRVideoSET dataset for interested readers. Percentage bitrate savings are not presented in this paper as, for many cases, the overlapping area is small, and thus the bitrate-saving percentage obtained for the compared codecs is unreliable. Based on the results presented in the Table~\ref{table:BDBRresults}, the following observation can be drawn:
\begin{enumerate}
    \item In terms of both quality metrics (PSNR and HDR-VQM), AV1 results in the best performance amongst all the four encoders considering individual gaming videos as well as the average over all eighteen video sequences. 
    \item The performance of the compared codec combinations is highly dependent on the video content. For some games, while the performance might vary a lot (e.g., 2 dB for 
    AV1 vs. VP9 for RUSH-P1 and RUSHP2), for some other games
    the performance is almost the same across both video sequences from the same game.
    \item The percentage quality gain among the compared codecs varies with the content, and no particular observation can be concluded.
\end{enumerate}

\subsubsection{Quality Fluctuation} \label{subsubsec:BitrateFluctuation}

\begin{figure*}[t]
    \centering
    \begin{adjustbox}{minipage=\textwidth,scale=1.0}
    \begin{subfigure}[b]{0.5\textwidth}
    \includegraphics[width=1.0\textwidth]{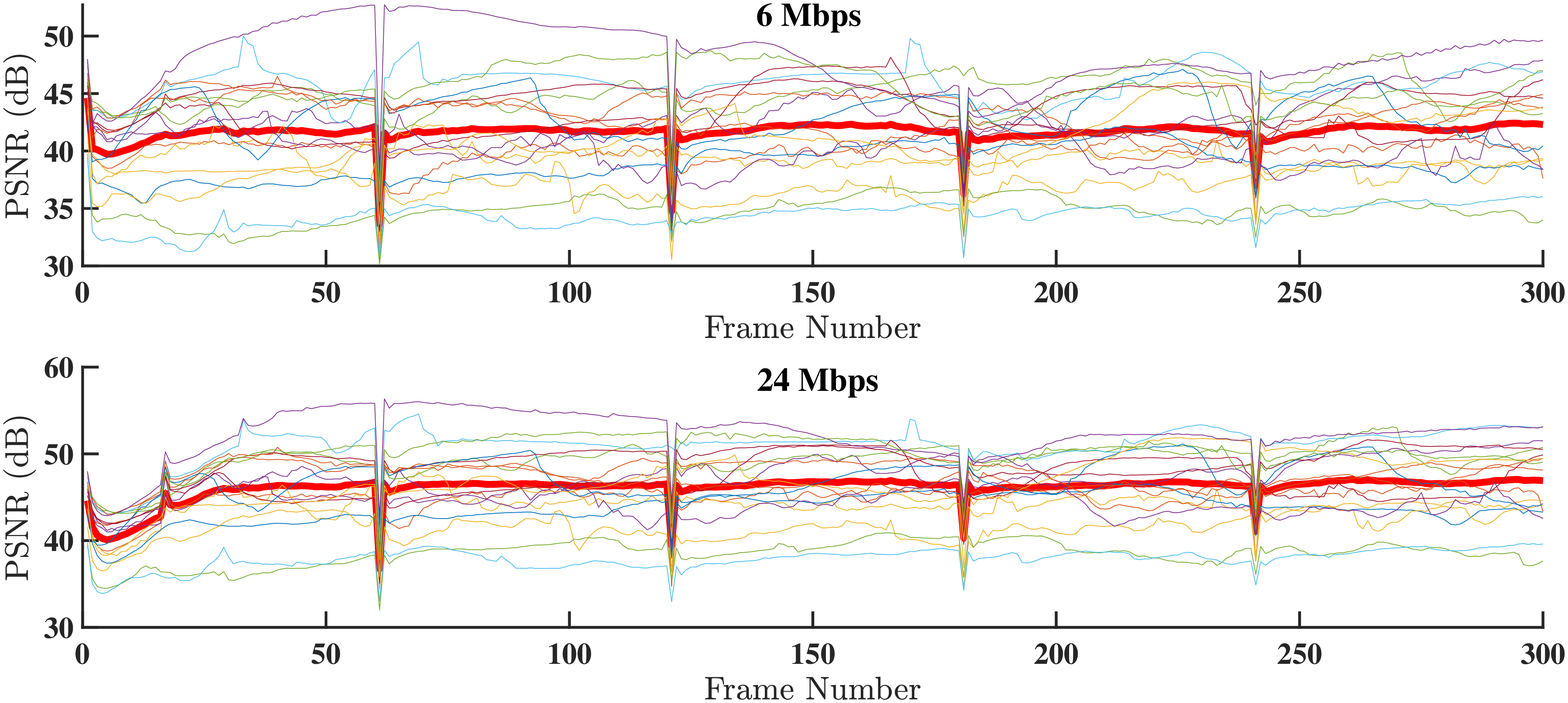} 
    \caption{Video sequences encoded using AV1 at 6 Mbps and 24 Mbps.}
    \label{fig:PerframeAV1}
    \end{subfigure}
    \begin{subfigure}[b]{0.5\textwidth}
    \includegraphics[width=1.0\textwidth]{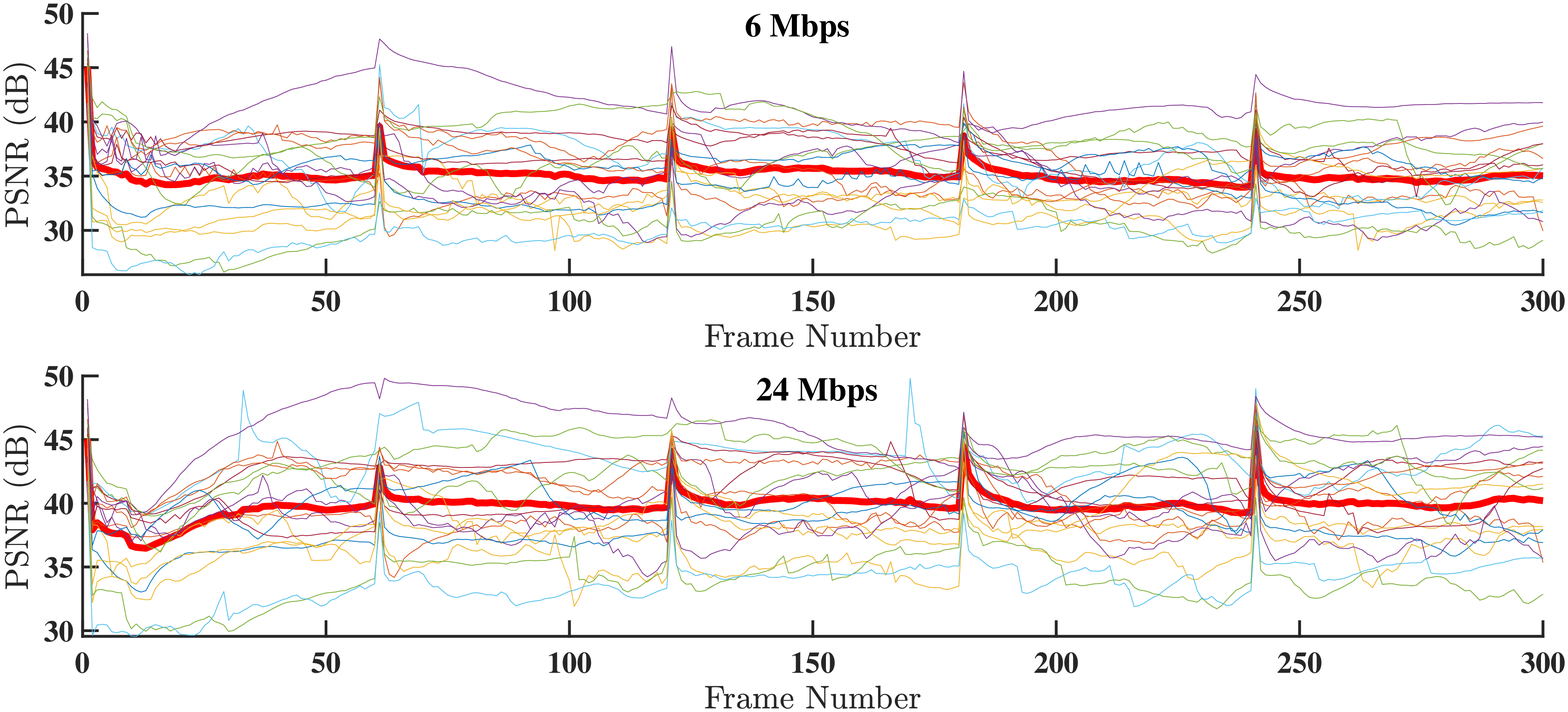}
    \caption{Video sequences encoded using VP9 at 6 Mbps and 24 Mbps.}
    \label{fig:PerframeVP9}
    \end{subfigure}
    \vspace{1cm}
    \begin{subfigure}[b]{0.5\textwidth}
    \includegraphics[width=1.0\textwidth]{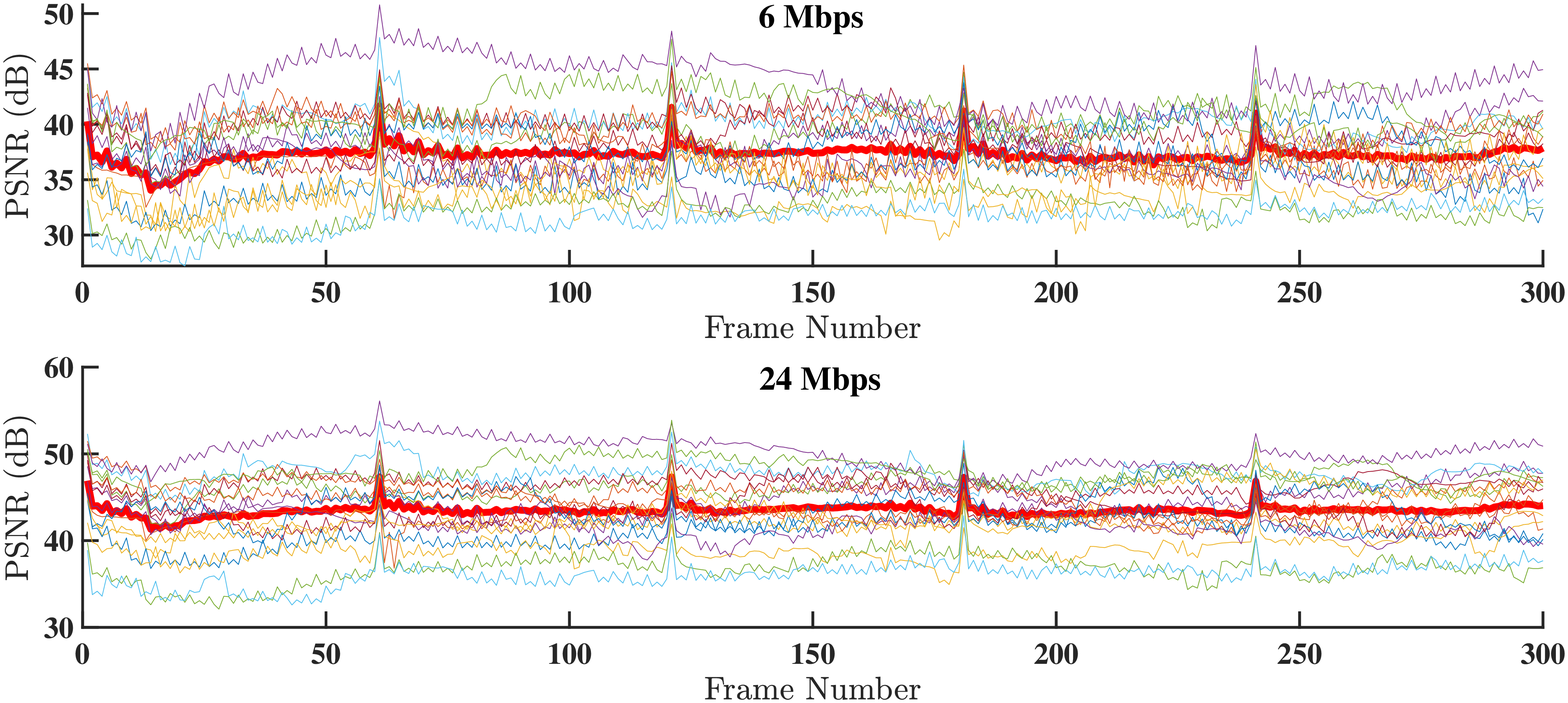} 
    \caption{Video sequences encoded using X264 at 6 Mbps and 24 Mbps.}
    \label{fig:PerframeAV1X264}
    \end{subfigure}
    \begin{subfigure}[b]{0.5\textwidth}
    \includegraphics[width=1.0\textwidth]{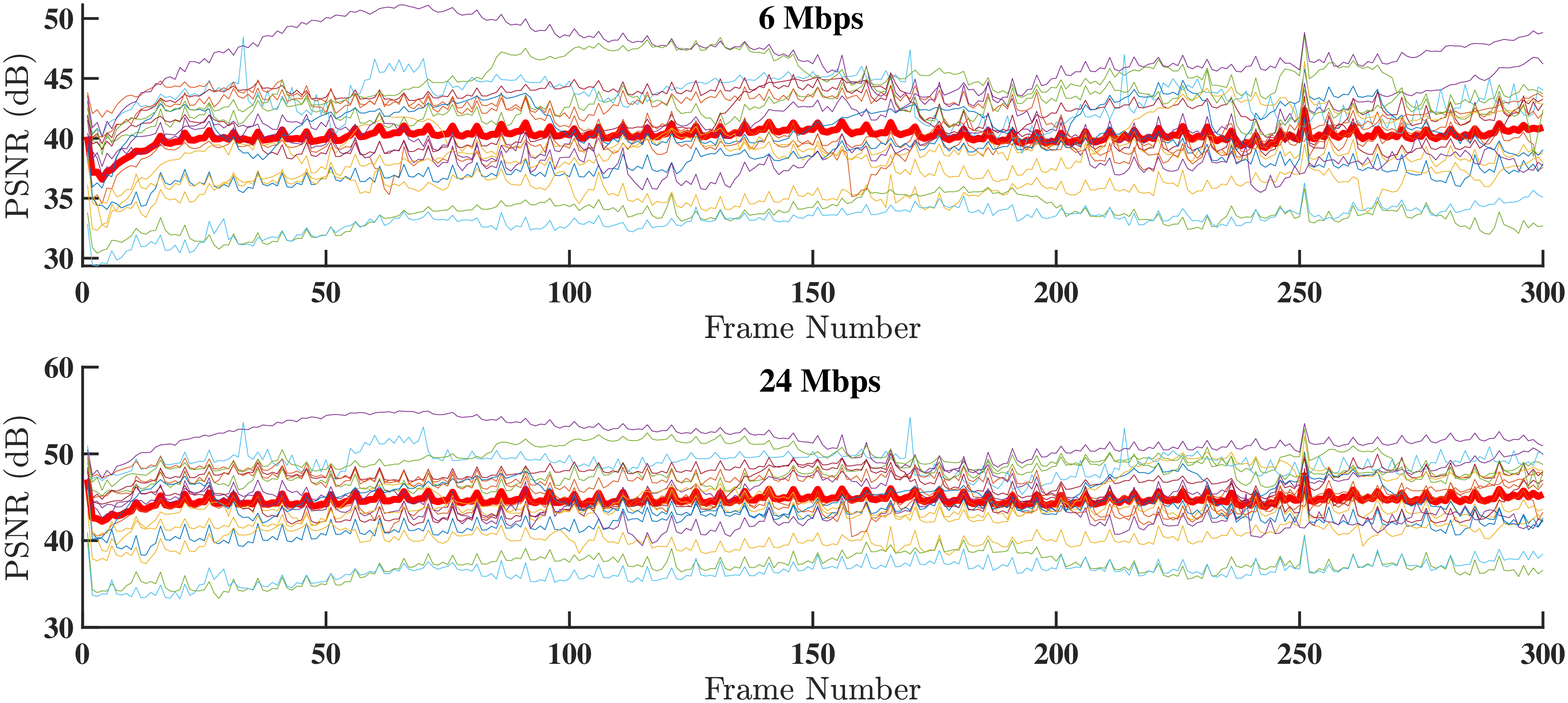}
    \caption{Video sequences encoded using X265 at 6 Mbps and 24 Mbps.}
    \label{fig:PerframeAV1X265}
    \end{subfigure}
    \end{adjustbox}
    \caption{Per frame PSNR scores of all 18 individual video sequences (coloured line curves) along with the average value curve (thick red line) considering all 18 video sequences encoded using AV1, VP9, X264, and X265 at 6 Mbps and 24 Mbps.}
    \label{fig:PerFrameAllCodecs}
\end{figure*}

While we reported earlier the results using average quality scores obtained for the whole video sequence, as discussed by the authors in \cite{Barman2019Survey}, time-varying quality can affect the end-user \ac{QoE}. Hence, we present here an analysis of the quality variation of the video using per-frame quality (PSNR) scores. In order to evaluate how much the quality varies over the duration of the video, we extracted the per-frame PSNR values for each encoded video sequence. Due to the reasons discussed earlier, gaming videos are encoded using \ac{CBR} mode of encoding as it is preferred to keep the bitrate at a desired, constant value. Hence, we investigate the fluctuation of the quality of the encoded video sequences using per-frame PSNR scores. Figure~\ref{fig:PerFrameAllCodecs} presents the plot of the per-frame PSNR scores for all the 18 video sequences encoded using the four codecs at two different bitrates: 6 Mbps and 24 Mbps. 
The thick red line curve corresponds to the average value considering the per-frame scores of all 18 video sequences. The plots for other bitrates (12 and 18 Mbps) are not presented here for brevity but show similar behavior as the presented two bitrates and are included in the dataset.
The per-frame score variation of the four codecs provides some interesting insights. It needs to be reminded that all videos are encoded using 2-second fixed \ac{GOP} at \ac{CBR}, which results in an I-frame at frame number 1, 61, 121, 181, and 241. Based on the results provided in Figure~\ref{fig:PerFrameAllCodecs}, we can draw the following observations:

\begin{enumerate}
    \item \label{i-frame} \textit{I-frame quality}: A look at Figure~\ref{fig:PerframeAV1} for AV1 indicated a huge dip in quality for I-frames compared to P and B frames, which is in contrast to the other codecs where there is a quality increase for all three other codecs. The drop in I-frame quality is also visually noticeable during video playback. 
    \item \textit{Drop in initial quality}: For all four encoders, we observe an initial dip in quality, more visible at higher bitrate value. 
    The dip in AV1 and VP9 is more pronounced while it is the least for the H.265. This indicates that the rate-control optimization in the case of AV1 and VP9 is more conservative initially while allocating 
    the
    required bits and then averages over the duration of the video. The similarity between AV1 and VP9 is not surprising as AV1 uses VP9 as the codebase.
    \item \textit{Overall frame quality variation}: Considering the
    encoding bitrate of 6 Mbps, H.265 results in the least quality variation over all the frames, followed by H.264, AV1, and VP9. At 
    the other three bitrates, the order from least to highest variation is H.265, H.264, VP9, and AV1.
\end{enumerate}

\begin{figure*}[t!]
\centering
\begin{center}
\includegraphics[width=1.0\linewidth]{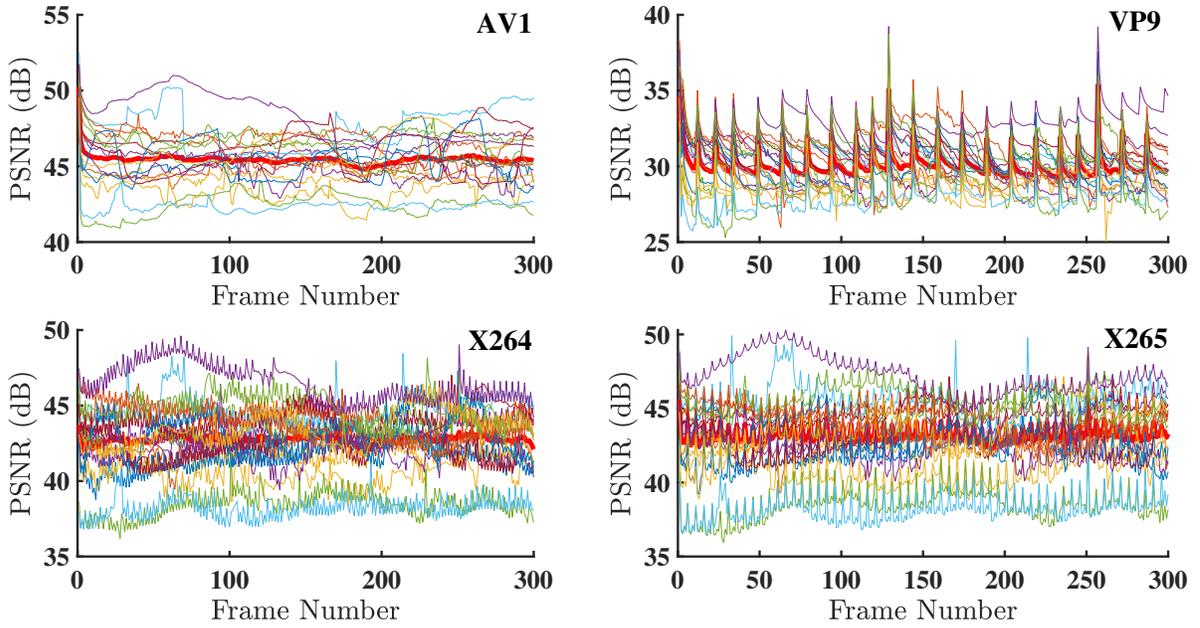}
\end{center}
\caption{Per frame PSNR scores of all 18 individual video sequences (coloured line curves) along with the average value curve (thick red line) considering all 18 video sequences encoded using AV1, VP9, X264, and X265 using CRF 23.}
\label{fig:PerframePSNRCRF}
\end{figure*}

In order to investigate 
if 
the reason behind the observation reported above in~\ref{i-frame})
is due to the choice of encoding mode (\ac{CBR} with closed \ac{GOP}) or 
other features of the encoder,
we encoded the videos using the \ac{CRF} mode of encoding, with no fixed \ac{GOP} size for all four codecs.
In Figure~\ref{fig:PerframePSNRCRF}, each coloured curve represents PSNR values over time for a sequence and the thick red line curve corresponds to the average value considering the per-frame scores of all 18 video sequences encoded using CRF of 23.

It can be observed that using the CRF mode of encoding, AV1 results in a much smoother frame quality across the frames than that observed earlier in Figure~\ref{fig:PerframeAV1}. The quality across frames for H.264 and H.265 results in frequent fluctuations in quality but of a quite low magnitude. On the other hand, VP9 sees quite big, regular change in frame PSNR quality. 

\subsubsection{Target vs. Actual Bitrate}

\begin{figure*}[t!]
\centering
\begin{center}
\includegraphics[width=1.0\linewidth]{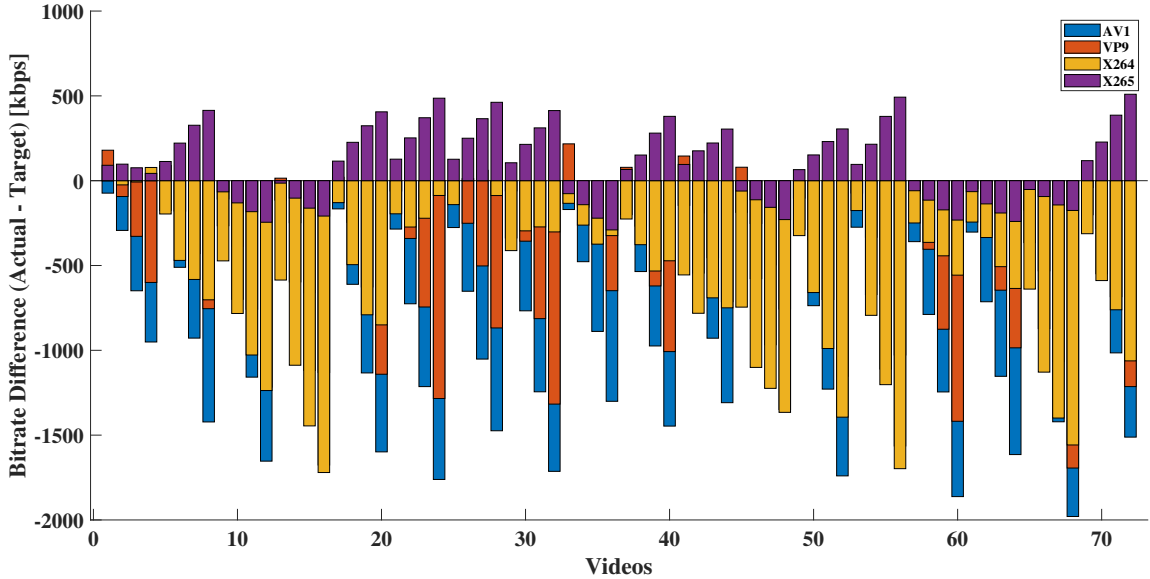}
\end{center}
\caption{Difference in actual and target bitrate for all four codecs for all 72 sequences.}
\label{fig:BitrateDiff}
\end{figure*}

Gaming video streaming is (usually) real-time in nature and 
the videos are encoded using \ac{CBR} mode of encoding, as discussed earlier in Section~\ref{sec:CodecComparison}. Due to the rate-distortion optimization in the codec, usually 
there is a difference between the target and the actual encoded bitrate.  For practical purposes, it is preferred that the actual bitrate value of the video is less than that of the target bitrate, as otherwise
it might lead to rebuffering issues at the client. Figure~\ref{fig:BitrateDiff} presents the bar plot for \textit{BitrateDiff} which 
is %
the difference in actual and target bitrate of the four codecs for all 72 video sequences. It can be observed that X.265 usually overshoots the target bitrate for most of the cases but results in the lowest average bitrate fluctuation ($93.28$ kbps). On the other hand, both X.264 and AV1 usually result in lower actual bitrate videos compared to the target bitrate - quite high in some cases (especially for higher bitrate encodes). On average, across all the 72 encoded video sequences, AV1, VP9, and X264 results in a bitrate
$873.14$, $491.88$, and $592.80$ kbps 
lower than the target one.
Further optimization of the rate-distortion optimization of these encoders for 1-pass \ac{CBR} mode of encoding might result in increased compression efficiency.

\subsection{Discussion}


 
For live streaming applications, a compromise is often required 
for the codec 
between speed (complexity) 
and
efficiency, as there is a real-time constraint for the encoding to be performed and 
streaming
to the end-user. While we did not report the encoding complexity in terms of encoding times, it must be noted that AV1 encoding 
complexity 
is many orders of magnitude higher than that of VP9, followed by H.265 and H.264. It is to be noted that the slower speed of FFmpeg's VP9 codec implementation library libvpx-vp9 is also reported in other studies, such as in \cite{Sharma2018VP9slow}.
The average PSNR gain 
of
X265 over X264 is $1.75 dB$, with it being just $0.5 dB$ 
for
some content such as PUBG. Similarly, the average gain obtained by AV1 over X265 is $1.58 dB$. Given the fact that the content 
is represented  
here 
in 10-bit and the multi-fold increase in content-encoding complexity of these newer codecs (AV1 and X265) over X264, the actual
benefit
of these newer codecs at the cost of such high encoding (and decoding) complexity using the discussed settings is questionable. Hence, as discussed in \cite{Ronca2019EncoderComplexity}, future video codec research must 
focus
on both compression efficiency and computational efficiency. In addition, due to the low
power requirements of devices such as smartphones, in addition to 
low encoding complexity, low decoding complexity is required. 
Netflix, for example, has already developed and started using an android AV1 decoder for its mobile streaming titles.

One of the major limitations of this work, as in many other codec comparison
works reviewed earlier, is that the analysis is limited to objective quality metrics. In our work, we used PSNR and HDR-VQM as the choice of quality metric for our codec comparison and frame quality variation for our 10-bit UHD-HDR gaming dataset. It is well known that PSNR does not correlate well with subjective scores \cite{Girod1993PSNRDoesNotWork}. Performance evaluation of the different quality metrics comparing their performance for video codecs for gaming and non-gaming 8-bit, \ac{SDR} content is presented in \cite{Barman2018NREvaluation} where it was found that the performance of quality metrics for gaming content
in some cases is different from that observed for non-gaming content. Hence, given that there has been no evaluation of the suitability of such HDR-VQM for \ac{HDR} gaming content, as well as considering the fact that recent studies such as 
\cite{Sugito2019HDR} have found that a simple \ac{SDR} metric can be applied directly to \ac{PQ} and \ac{HLG} encoded signals providing excellent results that surpass those of \ac{HDR} metrics, the performance gains reported here might be different from that observed subjectively. Also, in this work we limited our analysis to a single resolution,  hence limiting to only compression artefact 
while in many real-world applications, multiple resolution-bitrate pairs are used.
Codec compression performance results might be different when considering multiple resolution-bitrate pairs.


\section{Conclusion and Future Work} \label{sec:Conclusion}

In this paper, we presented GamingHDRVideoSET, the first 10-bit, UHD-HDR gaming video dataset, consisting of 18 reference video sequences
and 288 distorted video sequences obtained by encoding 
the references 
in four bitrates using four different video compression standards. Besides, per-frame and average PSNR and HDR-VQM quality metrics scores for the video sequences are provided. Reference video sequence analysis using \acl{DR} calculation, indicated 
\ac{DR} values for the reference video sequences are a bit lower
than 
for
professionally captured video sequences. 
The dataset presented here is not only the first considering gaming videos but also considering user-generated HDR content. Thus, the dataset can be used further for the design of better HDR signal processing, encoding (including HDR metadata), and display systems. While there exist no current applications which stream UHD-HDR gaming videos, with new players such as Stadia offering cloud gaming services capable of UHD-HDR gaming, it is time 
for 
the industry
to start offering such services in the very near future. Different considerations (profile, level, etc.) can be included in various codecs to further improve their compression efficiency for user generated HDR content. Such improvements can lead to faster availability and adoption of gaming HDR content. 

We also evaluated 
on the newly designed GamingHDRVideoSET dataset
the performance of H.264, H.265, VP9, and AV1, the four most famous and widely used video compression standards, using their implementation available in FFmpeg. 
Using PSNR and HDR-VQM as the quality metrics, we found that AV1 results in the highest quality, 
followed by H.265, H.264, and VP9. It was also observed that the gain in quality by AV1 (and even H.265) compared to H.264 is not so high when taking into consideration the  
tremendous increase in encoding complexity and hence
increased encoding duration. One of the reasons behind this might 
be 
the fact that our reference video sequences were obtained by decoding the already H.265/HEVC encoded (though at very high bitrate) gaming video sequences.
Hence, for practical applications, especially for Live video streaming, future video research, engineering, and standardization efforts will need to take into account both encoding as well as decoding complexity, as well as transcoding gains offered by various compression standards.

This work was the first attempt towards creating an open-source user-generated HDR video dataset. Our future work will include both objective and subjective evaluation of multiple resolution-bitrate encoded video sequences and the design of HDR gaming video quality metrics. In general, give the increasing proliferation of HDR gaming, there exist a plethora of future research opportunities, from HDR gaming video quality analysis to design of HDR VQM for gaming videos, to encoding and transcoding optimization. With the increasing demand for scalability to support such increased encoding complexity, blockchain-based distributed video streaming platforms such as Livepeer and Dlive can be an interesting alternative to the more prevalent centralized live streaming platforms.
In a nutshell, the future of UHD-HDR gaming is already here, with many exciting challenges and opportunities.

\section*{Acknowledgements}
Nabajeet Barman would like to thank Yasuko Sugito from NHK, Tokyo, Japan for her help with the MATLAB code for the calculation of Dynamic Range for the HDR videos.

\bibliographystyle{ieeetr}
\bibliography{ms.bib}

\begin{IEEEbiography}
[{\includegraphics[width=1in,height=1.25in,clip,keepaspectratio]{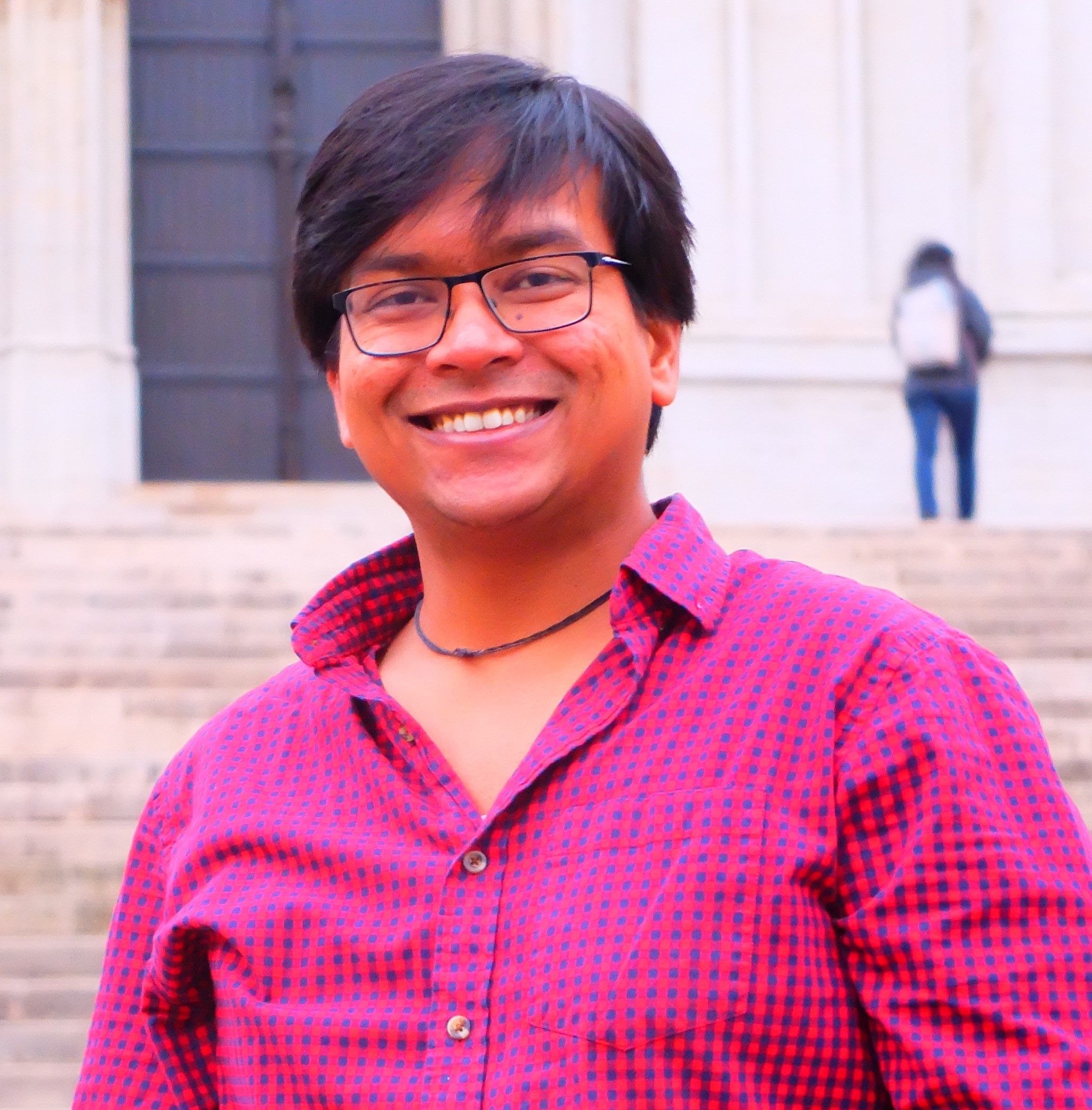}}]
{Nabajeet Barman} (Member, IEEE) received the B.Tech. degree in electronics engineering from National Institute of Technology, Surat, India, in 2011, the Masters degree in information technology, specializing in communication engineering and media technology from Universität Stuttgart, Germany, in 2014 and, MBA and Ph. D. degree from Kingston University London in 2019. Currently, he is a Lecturer in Applied Computer Science with Kingston University, London. From 2012-2015, he worked in different capacities across various industries including Bell Labs, Stuttgart, Germany after which he joined Kingston University as a Marie Curie Fellow with MSCA ITN QoE-Net from 2015 to 2018, and a Post-Doctoral Research Fellow from 2019-2020. He is a Board Member of Video Quality Expert Group (VQEG) as part of the Computer Graphics Imaging (CGI) group and is an active contributor in various ITU-T Standardization activities. His main research interests include multimedia communications, machine learning, data science and more recently distributed ledger technologies.
\end{IEEEbiography}
\vskip -2\baselineskip plus -1fil
\begin{IEEEbiography}[{\includegraphics[width=1in,height=1.25in,clip,keepaspectratio]{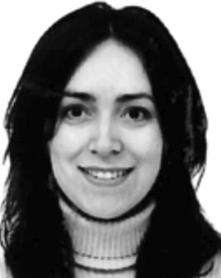}}]{Maria G. Martini} (Senior Member, IEEE) received the Laurea degree in electronic engineering (summa cum laude) from the University of Perugia, Italy, in 1998, and the Ph.D. degree in electronics and computer science from the University of Bologna, Italy, in 2002. She is a Professor with the Faculty of Science, Engineering and Computing, Kingston University London, where she also leads the Wireless Multimedia Networking Research Group. She has led the KU team in a number of national and international research projects, funded by the European Commission (e.g., OPTIMIX, CONCERTO, QoE-NET, and Qualinet), U.K. research councils, U.K. Technology Strategy Board/InnovateUK, and international industries. She has authored about 200 scientific articles, contributions to standardization groups (IEEE, ITU), and several patents on wireless video. Her research interests include QoE-driven wireless multimedia communications, decision theory, video quality assessment, and medical applications. She has been  Associate Editor of IEEE Signal Processing Magazine and IEEE Transactions on Multimedia,  the lead Guest Editor of the IEEE JOURNAL ON SELECTED AREAS IN COMMUNICATIONS special issue on “QoE-aware wireless multimedia systems” and  Guest Editor of journals such as the IEEE JOURNAL OF BIOMEDICAL AND HEALTH INFORMATICS, IEEE MULTIMEDIA,  the International Journal of Telemedicine and Applications. She has chaired/organized a number of conferences and workshops. She is part of international committees and expert groups, including the NetWorld2020 European Technology Platform Expert Advisory Group, the Video Quality Expert Group, and the IEEE Multimedia Communications Technical Committee, where she has served as the Vice-Chair from 2014 to 2016, as the Chair of the 3-D Rendering, Processing, and Communications Interest Group from 2012 to 2014, and as a Key Member of the QoE and multimedia streaming IG. She currently chairs the IEEE P3333.1.4 Standardization Working Group. She is an Expert Evaluator for the European Commission, EPSRC, and other national and international research funding bodies. 
\end{IEEEbiography}
\end{document}